\documentclass[pra,twocolumn,superscriptaddress]{revtex4-1}
\usepackage{amssymb}
\usepackage{amsmath}
\usepackage{amsfonts}
\usepackage{graphicx}
\usepackage{bm}
\usepackage{color}
\usepackage{multirow}
\usepackage{epstopdf}
\usepackage{hyperref}
\usepackage[normalem]{ulem}

\newcommand {\dchi} {$\partial \chi/\partial n$}
\newcommand {\dM} {$\partial M/\partial n$}
\newcommand {\ci} {$\chi^*$}
\newcommand {\cT} {$\chi_{\rm T}$}

\newcommand {\be} {\begin{equation}}
\newcommand {\ee} {\end{equation}}

\begin{document}

\author{V.~M.~Pudalov}
\email{pudalov@lebedev.ru}
\affiliation{P.\ N.\ Lebedev Physical Institute, 119991 Moscow, Russia}
\affiliation{National Research University Higher School of Economics, Moscow 101000,  Russia}

\author{A.~Yu.~Kuntsevich}
\affiliation{P.\ N.\ Lebedev Physical Institute, 119991 Moscow, Russia}

\author{M.\ E.\ Gershenson}
\affiliation{Serin Physics Lab, Rutgers University, Piscataway, NJ 08854, USA}

\author{I.~S.~Burmistrov}
\affiliation{L. D. Landau Institute for Theoretical Physics RAS, 119334 Moscow, Russia}
\affiliation{Laboratory for Condensed Matter Physics, National Research University Higher School of Economics, 101000 Moscow, Russia}
\affiliation{Institut f\"ur Theorie der Kondensierten Materie, Karlsruhe Institute of Technology, 76128 Karlsruhe, Germany} \affiliation{Institut f\"ur Nanotechnologie, Karlsruhe Institute of Technology, 76021 Karlsruhe, Germany}

\author{M.~Reznikov}
\affiliation{Solid State Institute, Technion, 3200003 Haifa, Israel}

\date{\today}

\title{
Probing spin susceptibility of a correlated two-dimensional electron system by transport and
magnetization measurements
}

\begin{abstract}
We report temperature and density dependences of the spin susceptibility
of strongly interacting electrons in Si inversion layers. We measured (i) the itinerant
electron susceptibility $\chi^*$ from the Shubnikov-de Haas oscillations in crossed magnetic fields and (ii) thermodynamic susceptibility \cT{} sensitive to all the electrons in the layer.  Both  \ci{} and \cT{} are strongly enhanced with lowering the electron density in the metallic phase. However, there is no sign of divergency of either quantity at the density of the metal-insulator transition $n_c$. Moreover, the value of \cT{}, which can be measured across the transition down to very low densities deep in the insulating phase,
increases with density at $n<n_c$, as expected. In the absence of magnetic  field, we found the temperature dependence of \ci{} to be consistent with Fermi-liquid-based predictions, and to be much weaker  than the power-law, predicted by  non-Fermi-liquid models.  We attribute a much stronger temperature dependence of \cT{} to localized spin droplets.
In  strong enough in-plane magnetic  field, we found the temperature dependence of \ci{} to be stronger than that expected for the Fermi liquid interaction corrections.

\end{abstract}

\maketitle
\section{Introduction}
Dilute two-dimensional  electron systems  (2DES) provide
unique opportunity for exploring the physics of strongly interacting charged
fermions, with spin properties of these systems being of particular interest.

Within the framework of the Landau theory of Fermi liquid (FL),  interacting electrons are described as a system
of quasiparticles with renormalized-by-interaction parameters,
such as  compressibility $\kappa^*$, effective mass  $m^*$,
spin susceptibility $\chi^*$, and $g$-factor $g^*$ \cite{landau, pines, landau_JETP_1956,
three_authors}.  These parameters are predicted to grow with lowering temperature and density, and their measurement is an important test of the theory. Strong renormalization of $\chi^*$ and $m^*$ was indeed observed in $^3$He and is well-explained within the framework of the Fermi liquid theory\,\cite{landau, pines}.

Disorder, especially in 2DES, greatly complicates the problem bringing into play another parameter,  the density $n_c$ of the apparent metal-insulator transition (MIT). This parameter happened to be very important: the spin susceptibility $\chi^*$ has been consistently found to grow strongly as the density $n$ is lowered towards $n_c$. Increase of $\chi^*$ relative to the non-renormalized Pauli susceptibility $\chi_P$ by a factor of 7
with decreasing electron density\,\cite{note1, ando_review}
was deduced from Shubnikov-de Haas (ShdH) oscillations in   tilted or crossed magnetic
fields\,\cite{okamoto_PRL_1999, gm, zhu_PRL_2003, granada, tutuc_PRL2002, clarke_Natphys_2008} and from analysis
of the in-plane magnetoresistance (MR) data\,\cite{shashkin_PRL_2001,
vitkalov_PRL_2001, sarachik_JPSJ_2003}.

Strong renormalization of $\chi^*$ with $n\rightarrow n_c$ raised an intriguing possibility of
magnetic instability\,\cite{finkelstein_SSR_1990} of a diluted disordered 2DES. As a result, much effort was invested in extracting \ci{}
as a function of {\em density}, typically at the lowest accessible temperature, with several papers claiming divergent \ci{} with $n\rightarrow n_c$ and $T\rightarrow 0$,
and therefore a quantum phase transition. Temperature dependence of \ci{}, which provides an important test for the theoretical models, was somehow overlooked.

In this paper we focus on the temperature dependence of the spin susceptibility obtained from  magneto-oscillatory transport (\ci{}) and from
thermodynamic (\cT{}) measurements of high-mobility dilute electron gas in Si inversion layers. We, (i) on the basis of both \ci{} and \cT{}
refute the claim for magnetic instability at $n=n_c$ , (ii) report temperature dependence of \ci{}, which happened to be weak and not
entirely consistent with available theories, and (iii) report thermodynamic susceptibility \cT{}, which is large and strongly temperature-dependent. Finally, we discuss the reason for the discrepancy between \ci{} and \cT{}.

\subsection{Density dependence of $\chi^*$: A brief overview}

In a clean system, ferromagnetic  instability originates solely from the combined effect of   electron-electron interactions
and  the Pauli principle. In metals the long range part of the Coulomb interaction is screened, whereas
the short range part leads to
strong correlations  in the electron liquid.
In clean metals this short range part of the interaction leads to ferromagnetic (Stoner) instability at sufficiently
strong interaction.
Critical  $r_s$  values for the expected instability  in a single-valley 2D system  varied from 13 to 20 in early calculations\,\cite{note1, ando_review, isihara_82}. According to them, the
ferromagnetic transition is likely to be of the first order with a
complete, rather than partial spin polarization.
More recent
numerical Monte Carlo calculations\,\cite{foulkes_rmp_2001,varsano_epl_2001, dharma-perrot_prl_2003}
predicted  a stable, fully polarized liquid  phase at $T = 0$ for $r_s > 25 -–26$,
before the Wigner crystallization occurs at around $r_s \approx 35-37$\,\cite{tanatar-ceperley_1989}.
The valley degree of freedom,   even twofold,  suppresses the tendency to spin polarization making the
fully spin-polarized state unstable\,\cite{marchi_prb_2009, attaccalite_prl_2002}.

Disorder, on the contrary, favors spin polarization. Within the Wigner-Mott model\,\cite{camjayi-dobro_natphys_2008, dobro-kotliar_WM_PRB_2012},
the effective mass  and thus  the spin susceptibility are predicted to diverge in the vicinity  of the MIT:  $\chi\propto (n-n_c)^{-1}$, though, according to numerical calculation\,\cite{Waintal10},
the system stays unpolarized in the accessible density range ($r_s<10$) at not-too-strong disorder ($1/k_Fl\lesssim 1$).

In this context, a strong in-plane field-induced MR in Si inversion layers \cite{simonian_PRL_1997, pud-MR_JETPL_1997} was interpreted in Refs.\,\cite{shashkin_PRL_2001, vitkalov_PRL_2001, sarachik_JPSJ_2003} as a  signature of ferromagnetic instability at, or very close to $n_c\approx 0.8\times10^{11}$cm$^{-2}$ in the studied samples, corresponding to $r_s \approx 9$.

\subsection{Theoretical studies  of the temperature dependence of $\chi^*(T)$: a brief overview}
\label{overview_chi(T)}

Within the FL theory,
Fukuyama\,\cite{fukuyama_1981} and Altshuler et al.\,\cite{altshuler82}
calculated interaction corrections to electron spin
susceptibility  for a 2DES system in the {\em diffusive regime}
 (low temperatures, $T\tau \ll 1$) to be:

\begin{equation}\label{FukAlt}
\frac{\Delta\chi^*(T)}{\chi_P}\sim -
\frac{1}{k_Fl}
\ln(T \tau).
\end{equation}
 Here  $\chi_P=2\mu_B^2 m_b/\pi$ stands for non-renormalized Pauli susceptibility of electrons with bulk effective mass $m_b$ and twofold valley degeneracy, $\tau$ is the momentum relaxation time, $k_F$ and $l$ denote the Fermi momentum and the mean free path, respectively.
Throughout the paper we use units with $\hbar=k_B=e=1$.
 Taking into account the valley multiplicity $n_v=2$ for (100)-Si, and introducing the relevant FL parameters (see Appendix \ref{App1}),
 this formula can be written as:
\be\label{Burmistrov}
\frac{\chi^*(T)-\chi_P^*}{\chi_P^*}=\frac{\delta\chi^*(T)}{\chi_P^*}\approx
\frac{2}{ \pi k_Fl}\frac{n_v F_0^\sigma}{(1+F_0^\sigma)} \ln(T \tau),
\ee
where {$\chi_P^*=n_v \mu_B^2 m^*/[\pi(1+F_0^\sigma)]$ is the spin susceptibility of the FL at $T=0$, which
includes FL-type renormalization in the absence of disorder,
$m^*$ is the  effective mass renormalized by the electron-electron interaction,  and $F_0^\sigma$ denotes the FL interaction
parameter in the particle-hole triplet channel.

 In the  {\em ballistic regime} (high temperatures, $T\tau\gg 1$)
the leading-order in $T$ correction to $\chi^*(T)$ for the 2D FL 
 was found to be proportional to the temperature\,\cite{chitov-millis,
chubukov-maslov_PRB_2003, galitski-das_PRB_2005, betouras-efremov_PRB_2005,
chubukov-maslov_PRL_2005, belitz97}:

\be\label{chi}
{\Delta \chi^*(T)} = {\chi_P^*}A^2 \frac{T}{4E_F}
\ee
where $E_F$  is the Fermi energy.  Within the theory of interaction quantum corrections   for a single-valley system $A$ is given in Ref.\,\cite{chubukov-maslov_PRL_2005} by:
$A = (m^*/m_b) \sum_{n=0}^{\infty} (-1)^n (2n+1)F_n^\sigma/(1+F_n^\sigma)$. Here $m_b$ denotes the non-renormalized band electron mass.
The particular expression for $A$  depends on the parameter range and the approximation used.

Within the interval $r_s=3 - 7$ of our measurements, $|F_0^\sigma|\approx 0.3 \div 0.5$ is large\,\cite{klimov_PRB_2008, clarke_Natphys_2008},
$n=0$ term in the series of harmonics dominates, and Eq.\,(\ref{chi}) acquires a simple form \,\cite{chubukov-maslov_PRL_2005, galitski-das_PRB_2005}:

\be\label{chi_Galitsky}
\frac{\Delta \chi^*(T)}{\chi_P^*} \approx \frac{T}{2E_F} G,
\ee
where $G\approx (m^*/m_b)/(1+F_0^\sigma) \sim 1$ varies from 1.6 to 2.7 as $r_s$ increases from 3 to 7. It was noted in Ref.\,\cite{chubukov-maslov_PRL_2005} that
prefactor $A$ in Eq.~(\ref{chi}) determines  also the slope of the linear-in-$T$ correction to the conductivity
$\Delta \sigma(T)$ of the 2D FL in the ballistic regime \cite{ZNA, valley_diffusion, alexkun-vv_PRB_2007}. The temperature dependence of the conductivity and magnetoconductivity
was carefully studied in Refs.\,\cite{klimov_PRB_2008, pudalov-aleiner_PRL_2003} and found to be consistent with the calculated interaction corrections
\cite{ZNA}. This agreement encouraged us to perform comparison of  $\Delta \chi^*(T)$ with the
predictions based on the theory of interaction corrections.

 We note that in the experimentally accessible  parameter range both  ballistic and diffusive interaction corrections
Eqs.\,(\ref{Burmistrov})-(\ref{chi_Galitsky}) are expected to be rather small: $\Delta\chi^*/\chi_P^* \lesssim 3\%$.
For  high-mobility Si-MOSFETs the accessible temperature range is set by $ 0.05 < T\tau < 5$; the upper  bound is determined by
the temperature $T^*\sim 10$K, at which $\tau_\varphi (T^*)$  becomes comparable with $\tau$ and  quantum coherence vanishes.

With temperature decrease the system  is anticipated to  enter the diffusive regime,
$T\tau \ll 1$. Within two-parameter renormalization group (RG) theory,
interactions become renormalized by the dimensionless parameter $\ln\left(l_\phi/l\right)=-\ln\left(T\tau\right)$,
 where $\l_\phi$ is the phase breaking length. Particularly, spin susceptibility   is predicted to grow
strongly as $T\rightarrow 0$  \cite{finkelstein84, castellani84, castellani_PRB_1998}:
$\chi^* \propto T^{-\zeta}$ with $\zeta <1$ \cite{fink_Science_2005}.

An important question is whether a
2D system of itinerant electrons undergoes spin ordering
and conventional FL model breaks down when the
interactions are strong
and $T\rightarrow 0$\, \cite{abrahams97, khodel_FC90}.
In particular, Khodel et al.\,\cite{khodel_FC90} predicted that the quasiparticle energy
spectrum $\varepsilon(k)$ flattens at the Fermi level as a result of interactions, and,
above a certain critical value $r_s\approx 7$, the 2D Fermi
surface (a circle) breaks into the nested rings
\cite{khodel_FC90}. This dispersion instability should  result in an essentially
non-Fermi-liquid behavior, which would lead  to strong
temperature dependence of the spin susceptibility $\chi^* \propto
T^{-2/3}$ \cite{clark_chi(T)_05, zverev_05}.

The mean field theory\,\cite{andreev-kamenev_prl_1998} suggests that a low-dimensional disordered system may
undergo a finite temperature  spin polarization at significantly weaker interaction strength than its clean counterpart.
The theory predicts divergence of the spin susceptibility
$\chi^*$  at a disorder-dependent temperature $T_c$, below which the system should become ferromagnetic.
This tendency toward ferromagnetic
transition originates from the effective enhancement of   interactions by diffusive dynamics of  electrons.

To summarize this brief overview, FL theory and numerical Monte-Carlo calculations
predict only a weak temperature dependence of $\chi^*$ and no ferromagnetic instability for a multivalley system.
By contrast, a number of other theories  \cite{finkelstein84, castellani84, castellani_PRB_1998, andreev-kamenev_prl_1998, clark_chi(T)_05, zverev_05}
predict divergence of $\chi^*$ in the vicinity of metal - insulator or topological transition in a 2D system.

\subsection{Experimental studies of the density and temperature dependence of $\chi$
and their interpretation}

On the experimental side,  we are not aware of any {\em direct measurements}
of the temperature dependence of $\chi^*$ for the itinerant  2D electrons by
transport techniques such as magnetotransport and quantum magnetooscillations.
Particularly, in  ShdH oscillation  measurements\,\cite{gm, shashkin_PRL_2003, zhu_PRL_2003}
no temperature dependence of the susceptibility was observed. Indeed, this dependence is very weak, as seen from our data, and therefore requires  high precision ShdH measurements in vector fields.

In Refs.\,\cite{anissimova_nphys_2007, knyazev_PRL_2008}  the temperature dependence of the weak field in-plane magnetoconductivity in  the vicinity of $n_c$  was found to behave as  $\Delta\sigma \propto \left(B^2/T^2\right)T^{-\varepsilon}$;  the factor $T^{-\varepsilon}$ was conjectured to originate from the renormalization of $\chi^*(T)$ in the spirit of  the two-parameter RG theory\,\cite{punnoose_PRL_2002}.
Clearly, this interpretation is model dependent. Based on such interpretation,
the strong growth of $\chi^*(T)$ was reported in Refs.~\cite{anissimova_nphys_2007, punnoose_PRB_2010, knyazev_PRL_2008}.
Later, in Refs.\,\cite{morgun_PRB_2016, pudalov-ICSM-2016}, this MR behavior was attributed
to semiclassical effects in the two-phase non-FL state. We also note that the magnetotransport measurements in tilted magnetic fields\,\cite{kuntsevich_PRB_2013}, performed
with the same or similar Si-MOS samples as those in Refs.\,\cite{anissimova_nphys_2007, punnoose_PRB_2010, knyazev_PRL_2008},
revealed that the in-plane MR cannot be described by electron-electron interaction correction only. Simple evidence is that the observed strong parallel
field MR quickly diminishes when the perpendicular field
component is applied on top of the parallel field \cite{kuntsevich_PRB_2013}. These inconsistencies question the assumption required for the  applicability of the RG approach.

As the density decreases and interactions increase,
a part of the itinerant electrons become localized already well above the critical density $n_c$\,\cite{teneh_PRL_2012}.
This tendency strengthens with density approaching $n_c$.
In contrast to the transport measurements, which are sensitive to the most conductive parts of a sample, thermodynamic measurements\,\cite{prus_PRB_2003, shashkin_PRL_2006,teneh_PRL_2012} probe the magnetization averaged over the {\em whole 2D  system}. Since the localized and mobile electrons
coexist and strongly interact, magnetic state of the localized electrons
may affect the transport properties.

Strongly temperature-dependent  $\partial M/\partial n \equiv -\partial \mu/\partial B$ was observed already in\,\cite{prus_PRB_2003}. However, contrary to the predictions of the  FL-theory in the ballistic regime \,\cite{chubukov-maslov_PRB_2003, galitski-das_PRB_2005, betouras-efremov_PRB_2005, chubukov-maslov_PRL_2005} for the Pauli-type susceptibility, the measured  thermodynamic paramagnetic susceptibility \cite{prus_PRB_2003} {\em grew} as temperature decreased. To account for the anomalous sign of the $d\chi/dT$ found in Ref.\,\cite{prus_PRB_2003}, Shekhter and Finkel'stein  considered
rescattering of pairs in the Cooper channel, which leads to anomalous temperature dependence $\Delta(1/\chi(T)) \propto -T $ in the ballistic regime
\cite{shekhter-fink_PNAS_2006, shekhter-fink_PRB_2006}.

Reference~\cite{prus_PRB_2003}  mostly discussed  magnetization in strong magnetic fields, $\mu_B B/k_BT\gg 1$.
In subsequent thermodynamic measurements\,\cite{teneh_PRL_2012} performed in weaker magnetic fields, the magnetization-per-electron $\partial M/\partial n$  was found to overshoot the Bohr magneton  at a magnetic field of $\mu_B B/ k_BT\approx  0.25$,
and thermodynamic susceptibility was found to grow  much faster,  as $\Delta\chi(T) \propto 1/T^2$.
The  low field data  was shown  to originate mainly from the collective localized spins whose contribution
greatly dominates over and masks the Pauli spin magnetization of the mobile electrons\,\cite{teneh_PRL_2012}.
This interpretation was supported by the theory~\cite{sushkov_PRB_2013}.

\section{Experimental}
Both  transport   and
thermodynamic magnetization measurements
were performed on (100)~Si-MOS samples from two different
wafers: ``small'' samples,
 $2.5\times 0.25$mm, with moderate peak mobility $\mu^{\rm peak}\approx 2.4\,{\rm m^2/Vs}$,
(Si3-10, Si6-14), and  ``large'' ones,
$5\times 0.8$mm
with high mobility $\mu^{\rm peak} \approx (3.2 - 3.4)\,{\rm m^2/Vs}$ at $T=0.3$\,K (Si-5, Si-15, Si-UW1, Si-UW2).
 All samples had the $190\pm 20$\,nm thick gate oxide thermally grown in dry oxygen,  atop of which an Al film (gate) was deposited by thermal evaporation.
For crossed-field transport measurements, we preferably used small-size samples  to ensure homogeneity over the sample area  of the  perpendicular  field  generated  by a small split-coil system\,\cite{crossed}.
For the thermodynamic magnetization measurements, we mainly used large-size samples  to increase the signal.

The carrier densities referred to throughout the paper have been found from the SdH oscillations and the Hall effect measurements
(more details are given in Appendix \ref{sec:App_density}).
The elastic scattering time $\tau$ shown in the figures  was measured following the approach
of Refs.~\cite{ZNA, klimov_PRB_2008, pudalov-aleiner_PRL_2003}. First, the temperature dependence of the conductivity
$\sigma$ was measured at zero field and relatively high temperatures, where $\sigma$  linearly depends on $T$. Then, the  $\sigma(T)$-dependence
was linearly extrapolated to $T = 0$ and $\tau$ was found from $\sigma(T=0)$ using the Drude formula.

\subsection{Spin susceptibility $\chi^*$ of itinerant electrons probed by quantum magnetooscillations}
We determined $\chi^*$ from the period and phase of
the beating patterns of  the SdH oscillations\,\cite{gm}.  In the crossed-field (or vector-field) technique, the in-plane component of the magnetic field
partially
spin-polarizes the electron system, whereas a weak
perpendicular component probes the density of electrons in the
split spin-subbands (for details, see Ref.\,\cite{crossed}).
The data was collected
over the temperature range $T=0.1 - 1$\,K
for electron densities  $n=(0.77 - 10)\times 10^{11}$cm$^{-2}$, corresponding to the dimensionless
interaction strength  $r_s= 9.5 - 2.6$,
respectively\,\cite{note1, ando_review}. An AC current 0.1 -- 1\,nA at frequency 13\,Hz was supplied
from a battery operated current source to reduce the electron overheating\,\cite{prus_PRL_2002}.

\begin{figure}
\includegraphics[width=200pt]{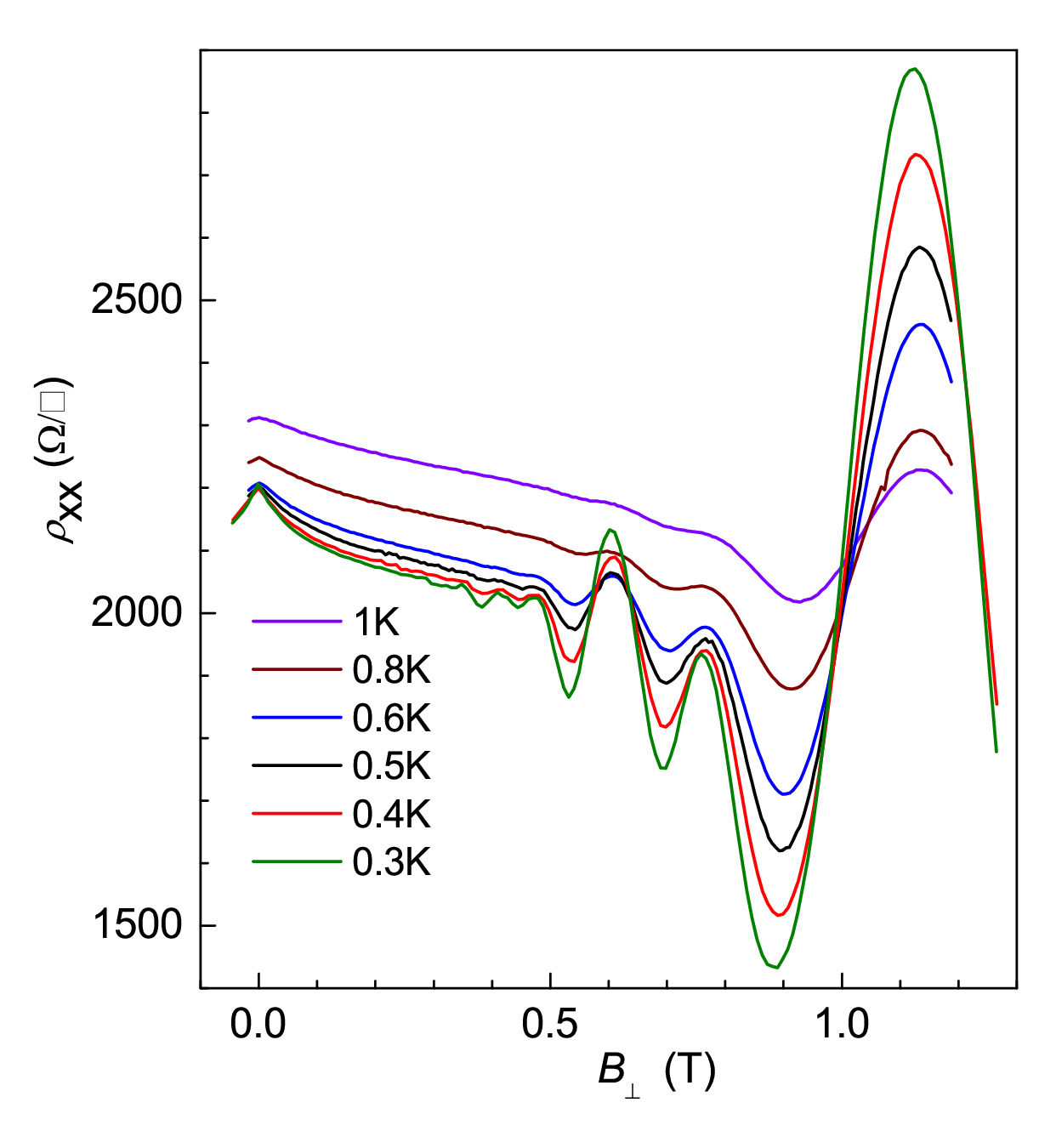}
\caption{(Color online)  Example of the $\rho_{xx}(B_\perp)$ data at different
temperatures.
$n\approx 2\times 10^{11}$cm$^{-2}$ and $B_\parallel=2.5$\,T.
``Large'' high-$\mu$ sample.}
\label{fig1}
\end{figure}

Figure\,\ref{fig1} shows an example of the raw $\rho_{xx}$ data  versus  perpendicular field $B_\perp$;
weak localization in fields $B_\perp< 0.1$T, smooth monotonic MR, and  ShdH oscillations are visible  for $B_\perp> 0.15$T.
The data in Fig.\,\ref{fig1} was taken with in-plane field $B_\parallel=2.5$\,T over the range $T=0.3-1$\,K.

\begin{figure}
\includegraphics[width=220pt]{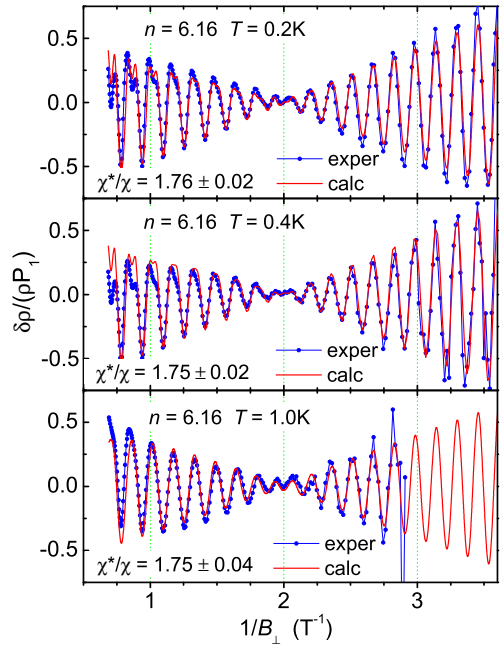}
\begin{minipage}{3.2in}
\caption{(Color online)
Examples of the normalized oscillatory magnetoresistance $\delta\rho_{xx}/(\rho_{xx} P_1)$ for high electron density
$n=6.16\times 10^{11}$cm$^{-2}$   measured at $B_\parallel =0.565$\,T and three temperatures.
Blue dots  are the data, blue connecting lines are guide to the eye, and red lines are fitting curves.  The extracted from fitting $\chi^*$ values
and their uncertainty are indicated in each panel.
A ``small size''sample.
 }
\label{fig2}
\end{minipage}
\end{figure}

To extract the oscillatory component,  we subtracted the second order polynomial \cite{gm} from the raw data (such as in Fig.\,\ref{fig1}).
When the amplitude of oscillations is small, the remaining oscillatory component $\Delta \rho_{xx}$ is well
described by the conventional Lifshits-Kosevich (LK) formula\,\cite{SdH, bychkov_SdH, isihara_86, pudalov_PRB_2014}.
Application of an in-plane magnetic field induces beating of the ShdH
oscillations\,\cite{gm};
examples of beating patterns  at high and low electron
densities are shown in Figs.\,\ref{fig2} and \,\ref{fig3}, respectively. For clarity the $\delta\rho_{xx}(B_\perp)$ data was normalized by
the amplitude  $P_1(B_\perp)$  of the first harmonic of the
oscillations\,\cite{gm, SdH, pudalov_PRB_2014} (for more detail, see Appendix \ref{sec:Appendix_2}).

One can see from Figs.\,\ref{fig2} and\,\ref{fig3}
that the oscillations can be precisely
fitted by the LK formula. Since  $B_\parallel$ is weak relative to the polarizing field,
and the interval of $B_\perp$ is narrow, there is no need in using empirical corrections to the LK formula\,\cite{pudalov_PRB_2014}.
We, therefore,  were able to fit the data with a single adjustable parameter, $\chi^*/\chi_P$.

In stronger magnetic fields, as  exemplified in
Figs.~\ref{fig2} and \ref{fig3},  at $B_\perp \gtrsim
1$\,T, the shape of the oscillations starts to deviate from the LK
formula.  This is caused by magnetic field dependent variations
of screening and level splitting due to the interlevel interaction, the latter is known to be  significantly enhanced in the quantum Hall effect regime\,\cite{pudalov_JETPL_1986, krav_PRB_1990, macdonald_PRB_1986}.
For this reason, below we analyze only  data obtained at $B_\perp\leq 1$\,T.

The procedure of finding  $\chi^*$ from the interference pattern
is straightforward and does not involve model assumptions; the
$\chi^*$ value is predominantly determined by
 the position of the
beats and by the phase
of the oscillations,  which sharply changes
by $\pi$ through the node. For this reason, the
uncertainty in the $\chi^*$ values is rather small, $\sim 0.5 - 4 \%$, which enabled us to detect
temperature variations of $\chi^*$.

In general, the accuracy decreases with temperature and reduction in $B_\parallel$,
therefore we were able to extract $\chi^*(T)$ only for
$T<1$\,K, where SdH oscillations are clearly resolved  in
weak $B_\perp$ fields\,\cite{gm, superstripes_2017, pudalov_MISM-2017}. Most sensitive to the value of $\chi^*$ is the node
position in the perpendicular fields $B_\perp \approx 0.5 -0.6$T. Therefore, we report $\chi^*$ values
obtained at such $B_\perp$ fields.
In the interval of low electron densities, $n=(1.1 \div 2.2)\times 10^{11}$\,cm$^{-2}$, the  main harmonic of the oscillations is suppressed
by the Zeeman factor (for more details, see Sec.~III\,D), and the spin splitting causes asymmetric nonharmonic oscillations. This  fact
favors  oscillation analysis and enables extracting $\chi^*$ with high precision in very weak and even in zero
$B_\parallel$ fields; the representative results are shown in Fig.\,\ref{fig4} (b).

We wish to stress that the deduced values of $\chi^*$ are independent of details of the fitting,
such as, e.g. field dependence of the oscillation amplitude.
The $\chi^*$  values rely  solely on the robust assumption  that an in-plane field causes Zeeman splitting
of the Landau levels $\hbar\omega_c(n+1/2)  \pm   g^*\mu_B B_{\rm total}/2$ which, in its turn,
leads to the phase shift  between  the  sequence of the oscillations produced by the ``spin-up''
and ``down'' electrons.
This method of finding $\chi^*$  also relies on the firm  fact
that the electron-electron interactions in the 2D system affects  neither the frequency nor the phase of the oscillations.

\begin{figure}
\includegraphics[width=225pt]{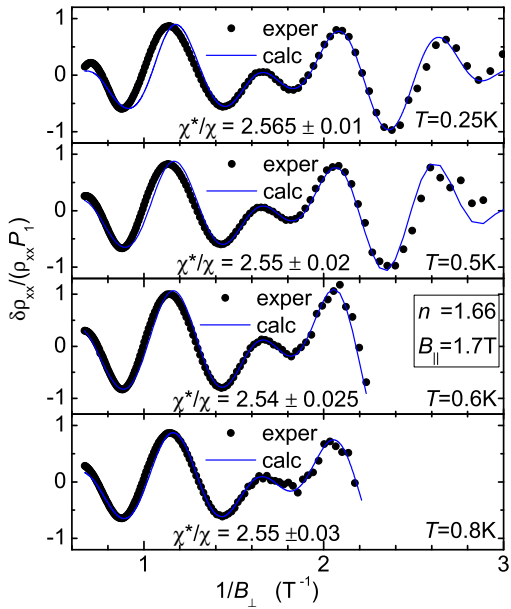}
\begin{minipage}{3.2in}
\caption{(Color online) Examples of the normalized oscillatory magnetoresistance
 $\delta\rho/(\rho P_1)$  for low electron density and four
different temperatures. Dots are the data and blue lines are the fitting curves. The nominal density  $n=1.66\times
10^{11}$cm$^{-2}$, $B_\parallel =1.7$\,T, a ``small'' sample.
}
\label{fig3}
\end{minipage}
\end{figure}

Figures\,\ref{fig4}(a)--4(d) show the obtained-in-this-way
temperature dependences of $\chi^*$
for three  carrier densities. In the explored
range of densities, the temperature dependence of $\chi^*$
is weak \cite{superstripes_2017}.  At even lower densities, close to $n_c$, $\chi^*$ is renormalized significantly.
Unfortunately, at such densities SdH oscillations can be observed  only
at the lowest temperatures,
and therefore temperature dependence of $\chi^*$ is experimentally inaccessible.
In the ``Discussion'' section, we compare the extracted $\chi^*(T)$ dependence with the theory.

\begin{figure}
\includegraphics[width=220pt]{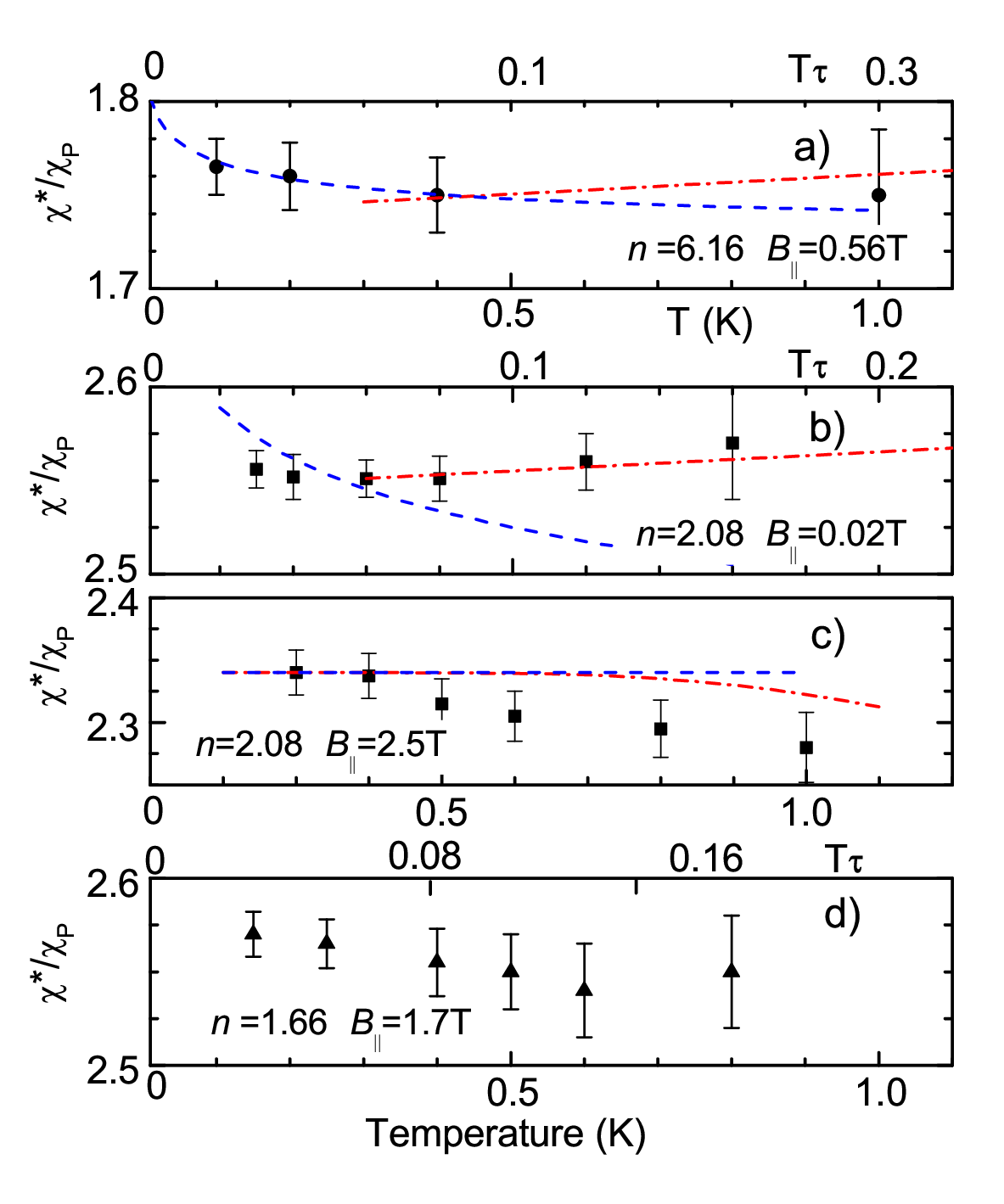}
\begin{minipage}{3.2in}
\caption{(Color online) Temperature dependence of $\chi^*$ for a ``small'' sample with  $n_c\approx 1\times 10^{11}$cm$^{-2}$
for three densities indicated in units of $10^{11}$cm$^{-2}$.
Upper axes show the temperature in units  of $T\tau$.   All the data corresponds to $B_\perp \approx 0.3-0.6$\,T;
the in-plane field component $B_\parallel$ is indicated  in each panel. Dashed and dash-dotted lines
show the  diffusive and ballistic corrections, respectively.
In calculations for panel (a), we used $r_s=3.35$, $F_0^\sigma=-0.3$, and for  (b) and (c),  $r_s=5.77$, $F_0^\sigma= -0.425$,\,\cite{klimov_PRB_2008}.
Diffusive and ballistic corrections were calculated using zero-field expressions
Eqs.\,(\ref{Burmistrov}) and (\ref{chi_Galitsky}) for panels (a) and (b), and finite-field Eqs.\,(A7) and (A8) for panel (c).}
\label{fig4}
\end{minipage}
\end{figure}

\subsection{Thermodynamic spin susceptibility \cT{}}
Thermodynamic susceptibility $\chi_{\rm T} =\left . dM/dB\right |_{B=0}$, where $M$ is the magnetization per unit area,
was obtained using the recharging technique\, \cite{prus_PRB_2003, reznikov_JETPL_2010, teneh_PRL_2012}, in which the chemical
potential response $\partial\mu/\partial B$ to a modulation of in-plane magnetic field is measured.
By virtue of a Maxwell relation,  it can be expressed as a derivative of $M$ with respect to the density:
$\partial\mu/\partial B=-\partial M /\partial n$. From the low field slope, $\partial \chi_{\rm T}/\partial n = -\partial^2 \mu/\partial B^2$ was extracted.

External magnetic field modulation with amplitude $\delta B(\omega)$ at a constant gate voltage leads to a modulation of the charge $\delta Q(\omega)$ given by:
\begin{equation}\label{dmudb}
\delta Q(\omega) = \frac {C(\omega)}{e} {\frac{\partial\mu}{\partial B}}\;\delta
B(\omega),
\end{equation}
\noindent
where $C(\omega)$ is the sample capacitance.  It is this charge modulation which is detected in the recharging technique.
Extracted from Eq.\,(\ref{dmudb}), ${\partial\mu}/{\partial B}$ was converted into ${\partial M}/{\partial n}$ using the Maxwell relation.
The details of this experimental technique have been provided in Refs.\,\cite{prus_PRB_2003, reznikov_JETPL_2010, teneh_PRL_2012}.
The thermodynamic susceptibility in principle can be obtained by integration of the low-field slope $d\chi_T/dn={\partial^2 M}/{\partial n \partial B}$ over density:

\be \chi(n,T)=\int\limits_0^n \partial \chi/\partial n(n',T)dn'
\label{int}
\ee

In practice, the lowest density $n_{L}(T)$ down to which the recharging technique works is set by the sample and contact resistances, which become large with lowering the density and temperature.
Importantly, it was realized\,\cite{reznikov_JETPL_2010} that the  technique can be used down to the densities substantially below $n_c$, even though the sample capacitance acquires imaginary part. Under such conditions, Eq.\,(\ref{dmudb}) should be used with the complex capacitance measured at the frequency of the magnetic field modulation.

Although we were able to measure \dchi{} down to the densities as low as $ 4\times 10^{10}\,{\rm cm^{-2}}$ at $T=1.7$\,K, still, to integrate we had to extrapolate \dchi{} to the interval $0<n<n_L(T)$. Figure\,\ref{Fig:chi} is obtained with \dchi{} constant and equal to its value at $n_L(T)$.  Taking instead \dchi$=0$ in this interval would shift the data in the figure by the value $\chi_{\rm T}(n_L)$; this would not lead to a qualitative change of the result.

In our earlier paper\,\cite{prus_PRB_2003}, in which the
thermodynamic method has been used for the first time, the measurements were performed down to $T=50$\,mK.
Treatment of the thermodynamic data in the current
paper differs from the previously published results  due to  different method of  integration of $\partial \chi/\partial n$ and $\partial M/\partial n$ over $n$: in Ref.~\cite{prus_PRB_2003} we integrated starting from a high density, assuming the magnetization at this density to be known. In the current paper we integrated $\partial M/\partial n$ from a low density. The minimal density $n_L(T)$ down to which
we can use this technique increases at low temperatures due to increasing resistance of both the 2DES and the contacts.
and no information can be obtained about magnetization at $n = n_c$. This complication is the primary reason to perform the current experiment at $^4$He temperatures rather than at $T\ll1$\,K. Besides this, we were interested in the low-field
behavior, $g\mu_BB \ll k_BT$. Measurements at much lower temperatures would require very weak magnetic fields, which in turn would lead to a small, difficult-to-measure recharging
current.  Note that the high-field (low-$T$) measurements  have been performed earlier in Refs.\,\cite{prus_PRB_2003, shashkin_PRL_2006}  and they  are not the goal of the current paper.

At low densities,  \cT{} grows linearly with $n$, which is expected for noninteracting electrons.
It reaches a maximum at some density $n(T)>n_c$ and then starts to decrease. We attribute this decrease
to melting of the spin droplets with density in the metallic phase. The maximum of $\partial M/\partial n$
and evolution of its position with temperature was discussed in detail in Ref.~\cite{teneh_PRL_2012}.

\begin{figure}
\includegraphics[width=230pt]{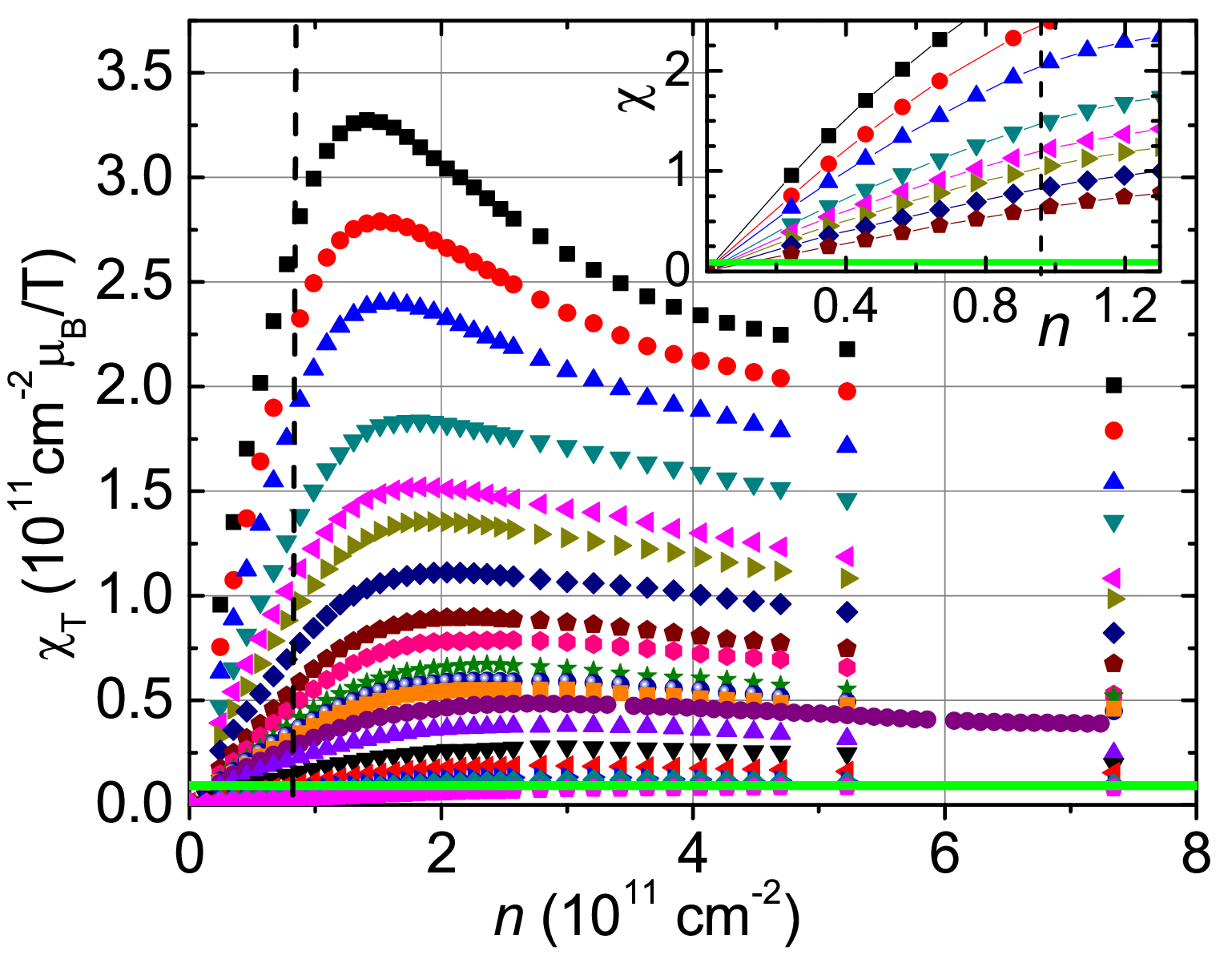}
\begin{minipage}{3.2in}
\caption{(Color online)  Thermodynamic spin susceptibility $\chi_{\rm T}$,  determined
from the chemical potential variations, versus the carrier density. Different curves correspond to temperatures
(from top to bottom)  1.7, 1.8, 2, 2.4, 2.7, 2.9, 3.1, 3.3, 3.5, 3.8, 4, 4.2, 4.6, 5.1, 5.7, 6.9, 8, 9.2, and 13K.
Green line shows the Pauli spin susceptibility $\chi_{P}$. The inset zooms in the  same data for the lowest densities
and eight temperatures. The dashed vertical line depicts the location of the critical density $n_c\approx 8\times 10^{10}\,{\rm cm^{-2}}$ for the studied sample.}
\label{Fig:chi}
\end{minipage}
\end{figure}

The thermodynamic susceptibility \cT{} at the lowest  temperature of 1.7\,K is a factor of 40 greater than the non-renormalized Pauli susceptibility.
As the temperature increases, \cT{} strongly decreases  as $T^{-\alpha}$ with $\alpha\sim 1-2$ depending on the density, see Fig.\,\ref{Fig:chiT}.
 This behavior contrasts sharply the
weak temperature dependence of the itinerant electrons' susceptibility $\chi^*$.
The temperature dependence of \cT{} is  much stronger than the theory\,\cite{shekhter-fink_PRB_2006, shekhter-fink_PNAS_2006} predicted for the itinerant electrons. Moreover, it is even stronger than the Curie law $\chi(T)\propto T^{-1}$ expected for noninteracting localized electrons. We, therefore, conclude that \cT{} is mainly due to easily polarized localized spin droplets\,\cite{teneh_PRL_2012}, and contribution of the itinerant electrons to \cT{} is negligibly small.
The fact that $\alpha >1$ in the $\chi_T(T)$ dependence points at the ferromagnetic type interaction, because random antiferromagnetic interaction  results in a spin susceptibility divergence slower than the Curie's low, with $\alpha <1$ \cite{sushkov_PRB_2013}.

\begin{figure}
\includegraphics[width=230pt]{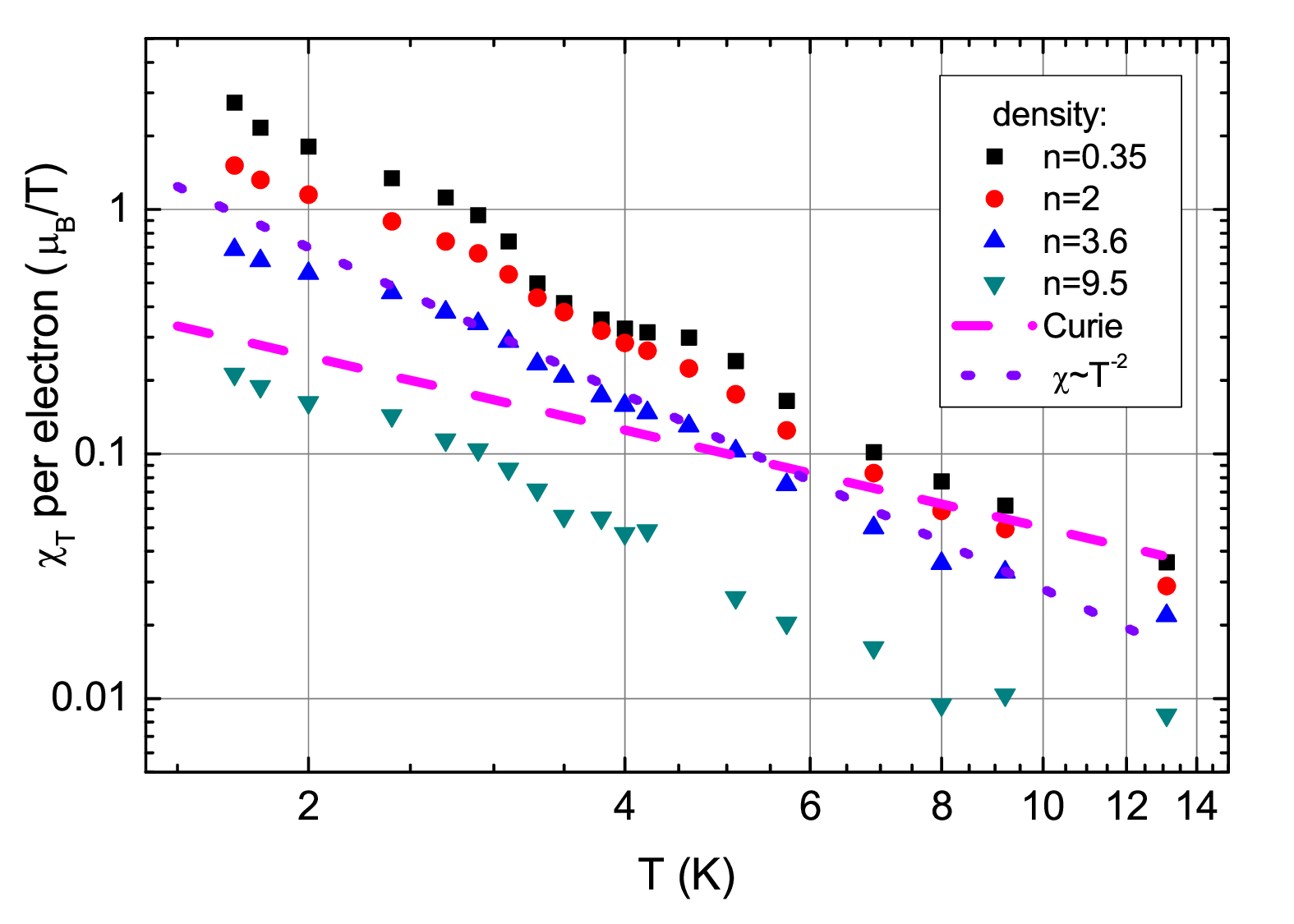}
\begin{minipage}{3.2in}
\caption{(Color online) Temperature dependence of \cT{} at different densities plotted for the same data
as in Fig.\,\ref{Fig:chi}. $\chi_{\rm T}\propto T^{-\alpha}$ with $\alpha\approx 2$. $\chi\propto T^{-2}$ and $\chi\propto T^{-1}$ (the Curie law) are shown for comparison. The fact that $\alpha>1$ means that the electrons, even at the lowest density, cannot be considered as noninteracting localized ones; this observation lead to the suggestion of the spin droplets in Ref.~\cite{teneh_PRL_2012}.}
\label{Fig:chiT}
\end{minipage}
\end{figure}

There is a contradiction between the observed increase of \cT{} faster then $1/T$ with decrease of the temperature and the upper bound for the magnetic moment: $M=\mu_Bn$.
Since \cT{} saturates\,\cite{reznikov_JETPL_2010,teneh_PRL_2012} at $B_c(T)\propto T$, such divergency would lead to a
divergent $M(B_c,T)=\chi_{\rm T} B_c\propto T^{1-\alpha}$. Therefore, we conclude that this divergency must be cut at a low temperature.  This conclusion is in accord with our early measurements\,\cite{prus_PRB_2003}, which
extended down to 50\,mK, and in which \dM{} was limited from above. Unfortunately, at such low temperatures the accessible range of densities starts from $n_c$,
and therefore one cannot make reasonable assumptions to carry the integration as in Eq.~(\ref{int}).

\subsection{Comparing $\chi^*(T)$ dependence for
the itinerant electrons with the interaction corrections}
To compare our data with  interaction  corrections\,\cite{altshuler82},
we plotted   in Figs.~\ref{fig4}(a)-4(c)
the diffusive and ballistic interaction corrections.
 The data is consistent with the  corrections at zero  and low $B_\parallel$-field, see panels (a) and (b).
In particular, there is a qualitative similarity between the data in Fig.\,\ref{fig4} and the ballistic corrections (see Fig.\,\ref{fig:diffusive_Burmistrov} of Appendix\,\ref{App1}).
Furthermore, the $\chi^*(T)$ temperature dependence  in panels (a) and (b)  exhibits a shallow minimum (thought this effect is weak and comparable with the error bars), expected for the ballistic-diffusive crossover\,\cite{ZNA}.
Indeed, as $T$ increases, the diffusive correction decreases  $\chi^*$  [see Eq.\,(\ref{Burmistrov})], whereas the ballistic -- increases\,\cite{betouras-efremov_PRB_2005}  [see Eq.\,(\ref{chi}), and for more details --  Appendix A].

The overall explored temperature range,
$T\tau \ll 1$, at first sight seems to belong to the diffusive interaction regime.
We note, however, that the conventional crossover criterion $T\tau = 1$ is  a qualitative estimate.
Quantitatively, the crossover temperature calculated  for transport \cite{ZNA} is $[(1+F_0^\sigma)/2\pi\tau]$,
i.e.  the crossover $T\tau$ value is expected to be  $\approx 0.1$ for our parameter range.
As for the spin susceptibility, the crossover regime has not been calculated until now.
Therefore,  we plotted   in Fig.\,\ref{fig4}, in addition to the diffusive correction Eq.\,(\ref{A7}), also the ballistic  one, using Eq.\,(\ref{eq:efremov}).
For weak field (Fig.\,\ref{fig4}\,(a)), we compare the data with
zero-field theoretical  result Eq.\,(\ref{Burmistrov}) rather than with Eq.\,(\ref{eq:lowB}), since the latter diffusive correction is applicable only above $T=(1+\gamma_2)g\mu_B B /2\pi\approx 0.5$\,K [see Fig.\,11(b)], i.e. in the region where predictions of Eq.\,(\ref{eq:B0}) [equivalent to Eq.\,(\ref{Burmistrov})] and\,(\ref{eq:lowB}) almost coincide.

Surprisingly, when a strong field, e.g. $B_\parallel = 2.5$\,T  which corresponds to  $g\mu_B B_\parallel\approx 3K$ [see Fig\,\ref{fig4}(c)],
is applied, the  $\chi^*(T)$  dependence does not vanish~\cite{altshuler82}. This observation
contradicts the interaction correction calculated for diffusive regime,
according to which the in-plane field $g\mu_B B_\parallel \gg T$ must  cut off  the $\chi^*(T)$-dependence, see Eq.\,(A7).
Instead, the temperature dependence of \ci{} changes sign, qualitatively similar to what is expected from  ballistic correction (see Appendix \ref{App1}).
Quantitatively, $\chi^*(T)$ reduces with temperature somewhat faster than the theory predicts (see Fig.\,\ref{fig4}\,c).
We note, however, that the uncertainty in the value of prefactor $G$ in Eq.\,(\ref{chi_Galitsky}) may also be $\sim 2$,
being related to the unknown details of the  electron-electron coupling,
and with  approximations used in deriving Eqs.\,(\ref{chi_Galitsky}), and~(\ref{eq:efremov}).

Importantly, since  $\Delta\chi^*(T)$ and $\Delta\sigma(T)$ stem from the same mechanism,  according to\,\cite{chubukov-maslov_PRL_2005}  they should  behave similarly with temperature.
Consistent with the theory\,\cite{ZNA}, weakening of the interaction correction to the conductivity $\Delta\sigma(T)$  by $B_\parallel$ field was reported for similar samples\,\cite{klimov_PRB_2008}, therefore the  anomalous $\Delta\chi^*(T)$ behavior in strong fields is very surprising.
A possible explanation might be that the diffusive-ballistic crossover for $\chi^*(T)$ is  shifted down in temperature  and corresponds to $T\tau=0.05 \div 0.1$ rather than unity.
It is also worth recalling  that we analyzed the ShdH oscillations beatings in the perpendicular field
0.5 $\div$ 0.6\,Tesla. A nonzero magnetic field is predicted\,\cite{betouras-efremov_PRB_2005} to shift the $\chi^*(T)$ minimum to $T= g\mu_B B$  in the ballistic regime. In the data, the minimum, indeed, shifts monotonically to higher temperature with increase of the total magnetic field, which occurs in sequence (b-a-d-c) in Fig.\,\ref{fig4}. However, the $\chi^*(T)$ minimum according to the ballistic theory should be located at about three times stronger fields.

The above listed inconsistencies lead us to the conclusion that the
temperature dependence of $\chi^*(T)$ in moderately strong magnetic fields, and maybe even to a lesser extent in small fields, is affected by another, stronger mechanism than the FL corrections. This  conclusion is supported by thermodynamic data, as will be shown below.

\subsection{Upper bound on $\chi^*$ as set by analysis of the
magnetooscillations}

Above we  reported the low-temperature $\chi^*$ obtained from the beats of the ShdH oscillation. Unfortunately, this technique cannot be used down to the critical density $n_c$ since the
in-plane magnetic field quickly drives the 2D system  into an insulator state\,\cite{pud_physicaB_1998, simonian_PRL_1997, pud-MR_JETPL_1997}.
Nevertheless, we still can use SdH oscillations to set the {\em limits} for \ci{}. For this, we
now  turn to the analysis of (1) the period, and (2) the  phase  (sign) of SdH oscillations just above $n_c$. The period  reflects the
degeneracy of the system,  which  would drop by a factor of two if the system becomes fully spin polarized and the Fermi energy doubles.
The phase depends  on the ratio between the spin and cyclotron splitting, and thus  carries information on the degree of spin polarization,
$(n_\uparrow-n_\downarrow)/n$.

We first look at the ShdH oscillations, with no $B_\parallel$ applied,  at a relatively high density, for which we know \ci{}
from the oscillation beats. We check that the oscillation frequency and phase  (sign) agree with the expectations, and then apply the same analysis
to even lower densities  for which ShdH beats could not be measured; this will set the upper limit for \ci{}.

The logic behind this approach can be understood by examining the schematic energy diagram in Fig.\,\ref{Fig:spectrum}.
We  presume Landau levels to be significantly broadened by scattering, which is definitely true close to $n_c$.  Therefore, in the following, we shall consider the valley splitting to be negligible. When the spin splitting   $g^*\mu_B B_{\rm total}$ is much smaller than the cyclotron energy $\hbar\omega_c$ and therefore unresolved,   each Landau level is fourfold-degenerate.
 The  period of the SdH oscillations  corresponds to $\nu=4$,  and $\rho_{xx}(\nu)$ minima are located at $\nu=4i$ (where $\nu= n h c/eB_\perp$ is the filling factor and $i$--an integer~\cite{gm, ando_review, SdH, isihara_86}).  This case is illustrated in Fig.\,\ref{Fig:spectrum}(a); other cases [Fig.\,\ref{Fig:spectrum}(b) and (c)] are described in the caption of Fig.\,\ref{Fig:spectrum}.

 Although the Mermin-Wagner theorem prohibits spontaneous spin polarization at $T> 0$,
one may expect to see a strong $\chi^*(T)$  growth as $T\rightarrow 0$ in the vicinity of a quantum ($T=0$) phase transition to the spin-polarized state. In such a case a half-period $\Delta\nu = 2$ of the SdH oscillations would be observed starting from  a certain density-dependent in-plane  magnetic field. This field would decrease to zero with the density approaching the quantum critical one $n_q$.
Indeed,  frequency doubling of the SdH oscillations induced by a strong external field  $B_\parallel \sim 2E_F^*/g^*\mu_B$
was experimentally observed by Vitkalov {\it et al.}~\cite{vitkalov_PRL_2000}.}

\begin{figure}[h]
\includegraphics[width=220pt]{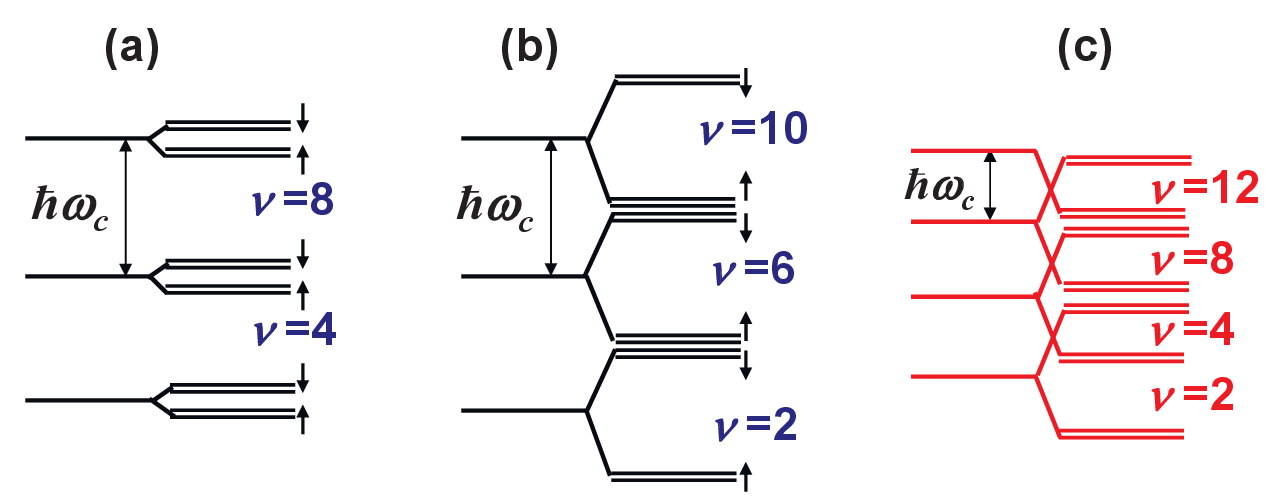}
\begin{minipage}{3.2in}
\caption{(Color online)  Evolution of the energy spectrum for the two-valley 2DES in (100) Si-MOS for various Zeeman splittings:
(a)  $E_Z \ll \hbar\omega_c$, spin splitting is unresolved and $B_\parallel$ field does not affect the oscillations amplitude. (b) $E_Z \lesssim \hbar \omega_c$, spin splitting dominates and increases with the field, causing the oscillations amplitude to grow with $B_\parallel$.  (c) $E_Z\approx 2\hbar\omega_c$\,--\,this situation does not occur in the studied Si-MOS samples.
Unresolved valley splitting is depicted with double lines. Filling factors denote the largest energy gaps.
}
\label{Fig:spectrum}
\end{minipage}
\end{figure}

\begin{figure}[tbh]
\includegraphics[width=230pt]{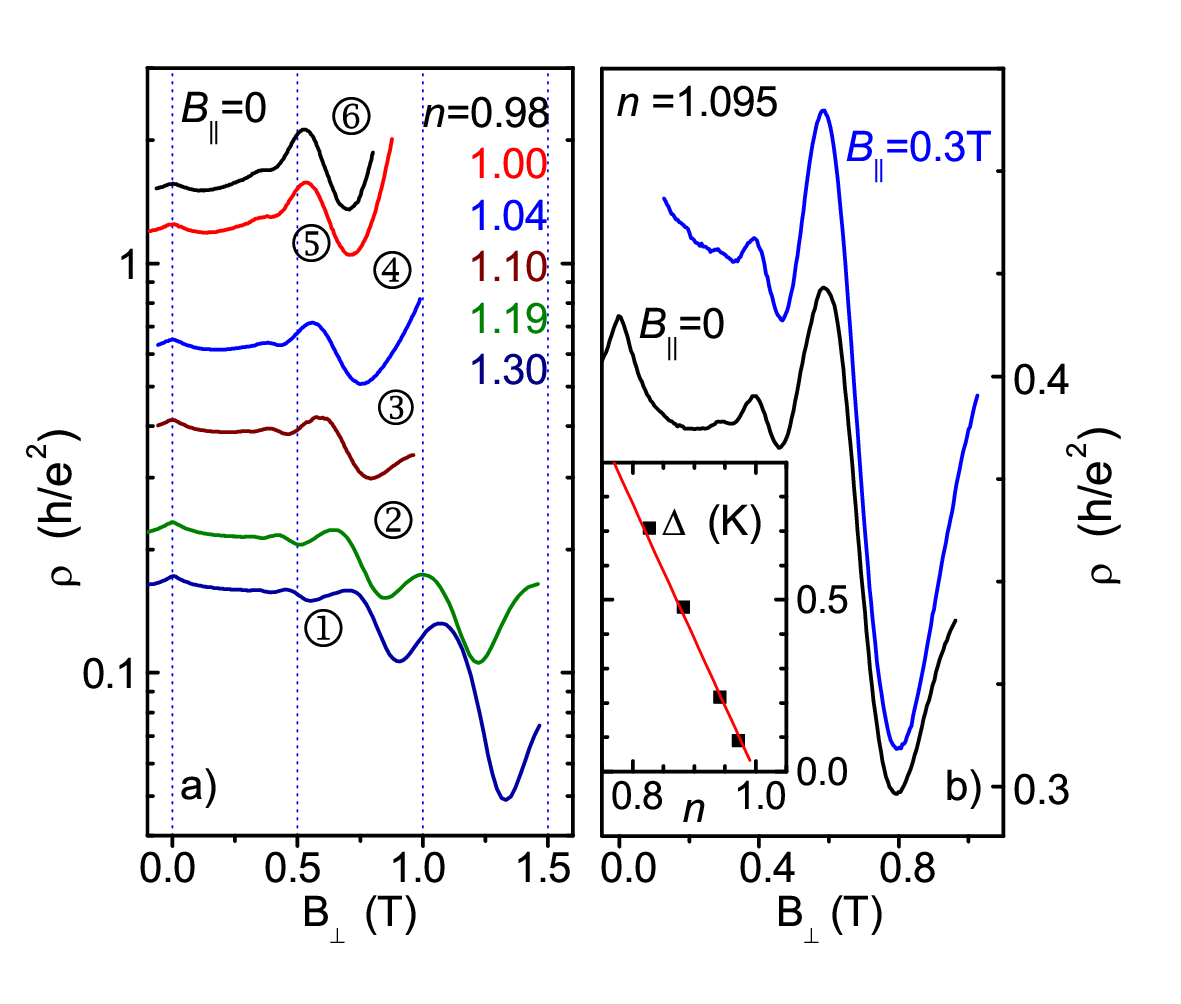}
\caption{(Color online) (a) ShdH oscillations for a  ``small'' sample
at six densities near $n_c\approx 1.0\times 10^{11}$cm$^{-2}$,   $T=0.2$\,K. Curves $3 - 6$ are terminated
at the onset of the large insulating peak in $\rho$
\protect\cite{pudalov_PRB_1992}. (b) Enhancement of oscillations with
$B_{\parallel}$.
 The inset in panel (b) illustrates the
determination of the critical density
from the activation energy $\Delta$, similar to \cite{pudalov_PRL_1993, pudalov_PRB_1992}. Densities are given in units of
$10^{11}$\,cm$^{-2}$.
}
\label{Fig:SdHnc}
\end{figure}

\begin{figure}[h]
\includegraphics[width=220pt]{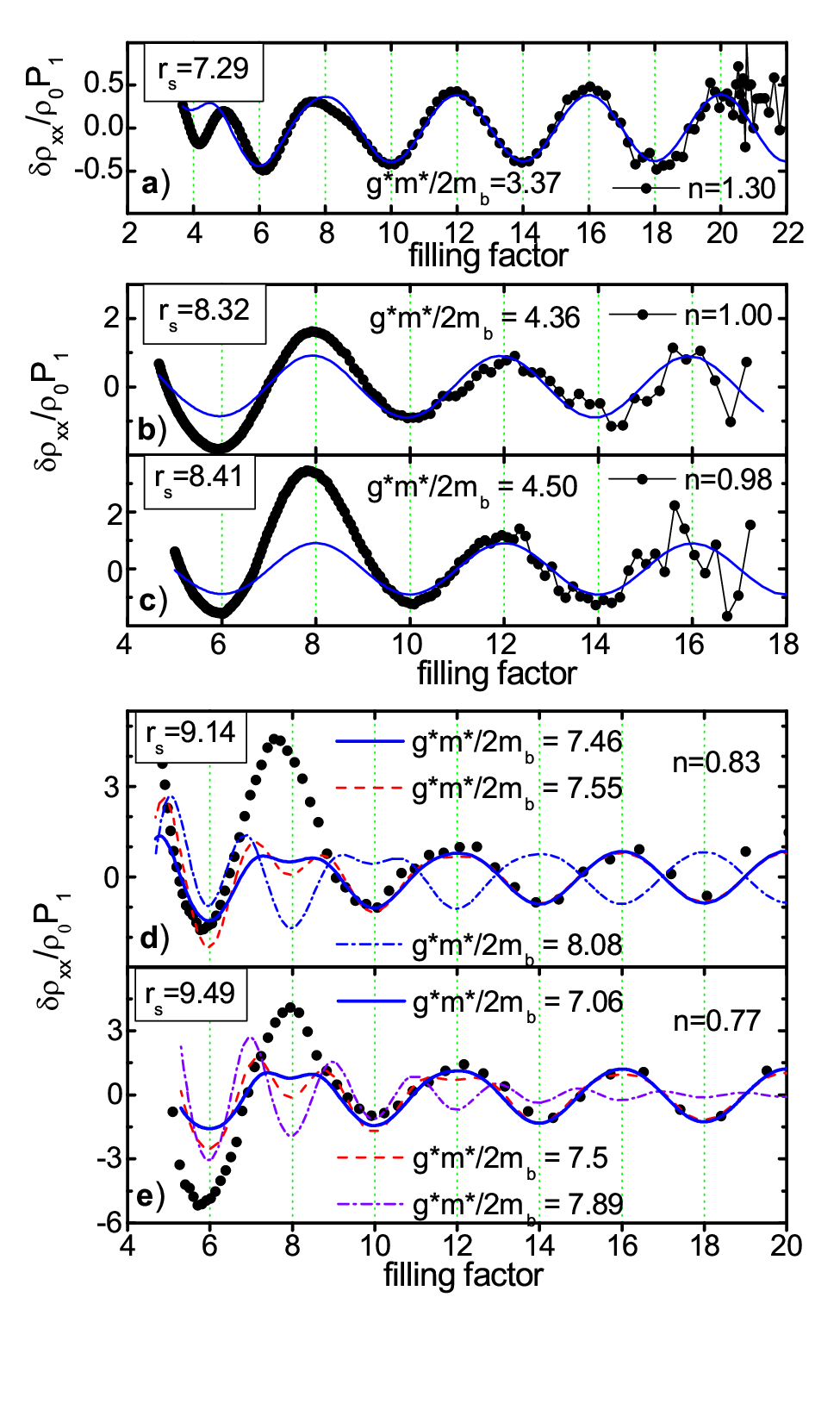}
\caption{(Color online)  Oscillatory component of the resistivity $\Delta \rho(\nu)/\rho_{0}P_1$ in $B_\parallel=0$.
(a)\,--\,(c) ``Small'' sample 
with  data {\em 1, 5} and {\em
6} from Fig.\,\ref{Fig:SdHnc}(a), $n_c=1.0\times 10^{11}$cm$^{-2}$.
(d), (e) ``Large''
sample
with $n_c=0.77\times 10^{11}$cm$^{-2}$.
Dots are the data,
fits are the lines\,\cite{gm} with parameters shown in the panels.
The temperature is 0.2\,K for panels (a - c) and 0.03\,K for
panels (d) and (e). The values of $n$ are in units of $10^{11}$\,cm$^{-2}$.
The red dashed and blue dashed-dotted lines in panels (d) and (e) show
what happens to the oscillation shape if one uses  inappropriate (increased) values of the fitting parameter $(g^*m^*/2m_e)\propto (\chi^*/\chi_P)$.
}
\label{Fig:9}
\end{figure}

Using this approach we now want to test the possibility of  anomalous  growth of $\chi^*$  in   a low  or zero  in-plane field with $n\rightarrow n_c$ and show that, if it exists, $n_q$ should be well below $n_c$.
Typical ShdH oscillations of $\rho_{xx}$  for very low densities, close to $n_c$,  are shown in Figs.\,\ref{Fig:SdHnc} and \ref{Fig:9}  as a
function of $B_\perp$. Due to high electron mobility, oscillations
are detected  in $B_\perp$ fields as low as 0.14\,T  for densities down to $n_c$ (which corresponds to $r_s\approx 9$).
The shape of these oscillations cannot be described with LK formula since (i) the electron-electron interactions at such densities are important \cite{pudalov_PRB_2014},
 (ii) close to $n_c$ a 2DEG is at the verge of the hopping  regime\,\cite{pudalov_PRB_1992, pudalov_PRL_1993}, and
(iii) at  $B_\perp\gtrsim 1$T the reentrant QHE-insulator transition quickly develops, preventing analysis of oscillations\,\cite{dio_physLett_1990, pudalov_PRB_1992}.
The parameter space where oscillations can be analyzed in the vicinity of $n_c$ is shown in  Appendix\,\ref{App3}.
We note that, unless the 2D system is fully polarized, the period of
oscillations is robust, since it depends only on the Landau level degeneracy,  which is not affected by interactions.

 For low densities  presented in Fig.\,\ref{Fig:SdHnc},  Zeeman splitting  is comparable to the cyclotron one
[see Fig.\,\ref{Fig:spectrum}(b)], and, therefore,  the $\rho(B_\perp)$ minima
in Fig.\,\ref{Fig:SdHnc}(a) are
due to the spin  gaps\,\cite{gm, okamoto_PRL_1999, termination94, krav_SSC_2000}.
Figure\,\ref{Fig:SdHnc}(b) shows
that the magnitude of the oscillations increases with in-plane
magnetic field $B_{\parallel}$, which enhances only spin splitting. The increase in the magnitude confirms that the ratio of
the Zeeman energy $E_Z=g^*\mu_B B_{\rm total}$ to the cyclotron energy
$\hbar \omega_c$ lies within the interval $1/2 <
E_Z/\hbar\omega_c < 1$ [see Fig.\,\ref{Fig:spectrum}(b)].
This estimation  is obtained from
the oscillation phase in the density range $r_s= 6.3$ -- 8.2; it  is in good agreement  with our earlier  quantitative measurements
of $\chi^*(n)/\chi_b \approx 5 - 6$ \cite{gm}   from the beating pattern of the ShdH oscillations in low crossed fields.

Curve {\em 5} in
Fig.\,\ref{Fig:SdHnc}(a) corresponds to the density $n=n_c$.
The latter value was determined by extrapolating the density dependence of the
activation energy $\Delta(n)$  for the exponential temperature dependences $\rho
\propto \exp(\Delta/T)$ in the insulating regime to zero\,\cite{pudalov_PRL_1993, pudalov_PRB_1992, pudalov_ECRYS_2002};
this procedure is illustrated in the inset to Fig.\,\ref{Fig:SdHnc}(b).

To clearly illustrate the phase  (sign) and period of ShdH oscillations in the low-field range,
we present in Fig.\,\ref{Fig:9} the  $\Delta
\rho_{xx}/ \rho_{0}P_1$  data obtained at $B_\parallel=0$ as a function of the Landau level filling,
$\nu=nh/eB_\perp$; the  data  is
 normalized by the calculated amplitude of the first SdH harmonic  $P_1(B_\perp)$  \cite{SdH, gm}.
In calculating  $P_1$, we used the values of  $g^*\propto h\chi^*(n) /m^*$ and $m^*(n)$ measured in Refs.\,\cite{gm, klimov_PRB_2008}.
 The Dingle temperature, $T_D$, [see Eq.~(B1)] was adjusted to match damping of the measured oscillations in  weak fields.

In strong fields, the oscillations amplitude is not described by LK formula, and is enhanced due to the factors  mentioned above.
Nevertheless, even in this case we obtain a reliable estimation of $\chi^*$, since  we rely solely on the  oscillation period and
phase, rather than the oscillations amplitude.

As seen from Fig.~9, down to the critical density for both samples, $n\approx   1\times 10^{11}$cm$^{-2}$ for the  small one (panels a-b-c) and
 $n=0.77\times 10^{11}$cm$^{-2}$ for the large one (panels d-e), the period of SdH oscillations in  weak $B_\perp$ fields  (large fillings) remains equal to
 $\Delta\nu=4$ providing evidence for  the  system to be unpolarized.
To illustrate this even further, we show in panels (d) and (e) a simulation with  $\chi^*$ increased by 6\%  (dashed line)
and 12\% (dash-dotted line); the appearance of the second harmonic and the change of oscillation phase by $\pi$ that would be caused
by such an increase is clearly visible.
From this we conclude  that $\chi^*$ value cannot be increased even by a relatively small factor.

In the analysis of ShdH oscillations it is important to limit the field range to $B_\perp \leq
1\,$T in order to avoid (i) {\em the magnetic-field-induced spin
polarization} and (ii)  {\em the emerging reentrant quantum Hall effect-to-insulator
transitions} \cite{dio_physLett_1990, pudalov_PRB_1992}.  This limitation  is violated at about
$\nu =5$:  a  shift of the minimum at $\nu\approx 4$ in Fig.\,\ref{Fig:9}(a)
is caused by  reason (i), i.e. by partial lifting of the spin degeneracy  in perpendicular
field $B_\perp \approx 1.3\,T$.
In Figs.\,\ref{Fig:9}(b) - 9(e) this limitation  is also violated for
$\nu<10$   and reason (ii)  may account for the excessive  oscillation
amplitude (for more details see Appendix \ref{App3}).

We  now  consider the phase  (sign) of  the oscillations.
The minima of
$\Delta \rho$ in Fig.~9 are located at $\nu=6, 10, 14, 18$, i.e.  at $\nu=(4i-2)$,
in contrast to  $\nu =4, 8, 12, 16$, i.e. $\nu=4i$, as observed for higher
densities. In other words, the  phase of the oscillations at low densities is reversed.  The position of the oscillations minima
is  consistent with   earlier  studies\,\cite{termination94, krav_SSC_2000, pudalov_PRB_1992},  with
values of $\chi^*$ measured from the oscillations beating \cite{gm}, and with the above analysis of the oscillation period (Fig.\,\ref{Fig:9}).
Schematically the energy spectrum for such a situation, $E_Z\gtrsim\hbar\omega_c^*$, is shown  in Fig.\,\ref{Fig:spectrum}(b).
The  oscillations phase changes by $\pi$ (i.e., sign changes)
due to the Zeeman factor
$\cos(\pi E_Z/\hbar\omega_c)$  when $E_Z$ exceeds $\hbar\omega_c^*/2$, i.e.  when  $\chi^*/\chi_P$ becomes equal to 1/2,
which occurs at $r_s > 6.3$ \cite{gm, SdH}, as schematically illustrated in Figs.~\ref{Fig:spectrum}(a) and 7(b).
The sign change is fully consistent with other observations (see, e.g.,  Refs.~\cite{termination94,  krav_SSC_2000}).

Since the phase of the SdH oscillations is determined by the ratio of
the Zeeman to cyclotron splitting \cite{SdH, bychkov_SdH}:

\begin{equation}
\cos \left( \pi\frac{g^*\mu_B B}{\hbar\omega_c^*}    \right)= \cos \left(\pi\frac{\chi^*}{\chi_P}\frac{m_b}{m_e} \right)
\end{equation}
it was concluded in Ref.\,\cite{spin-polarized} that,  to have  $\pi$ phase shift  in the range
$10 > r_s > 6$, the spin susceptibility $\chi^*$ must obey the following inequality:
$2.6 =m_e/2m_b < \chi^*/\chi_P <3m_e/2m_b=7.9$.
This inequality, together with a vanishingly small $\chi^*(T)$-dependence sets constrains
on the value of $\chi^*$ at the MIT critical density.

 The  above analysis of the period and phase of oscillations provides  strong evidence for
the absence of complete spontaneous spin/valley polarization of the itinerant electrons
at the sample-dependent MIT critical density  $n \geq 0.98\times 10^{11}$cm$^{-2}$
(for a ``small'' sample) and $n \geq 0.77\times 10^{11}$cm$^{-2}$ (for a ``large'' sample).

Another scenario, a potential divergence of
 $\chi^*$ and  $m^*$ for itinerant electrons
at a sample-independent density
$n_0$ was explored in Ref.~\cite{spin-polarized} by scaling the $\chi^*(n-n_0)$ dependence.
It was found that the $\chi^*(n) \propto g^*m^*$
data for both samples (the ``large'' and ``small'' ones) obey a common critical
dependence only if we choose $n_0 < 0.53\times 10^{11}$cm$^{-2}$, the unrealistically low value that belongs to the insulating regime.

This conclusion differs from the one suggested in Refs.\,\cite{shashkin_PRL_2001, vitkalov_PRL_2001}.
A possible reason for this disagreement is magnetization nonlinearity: strong magnetic field drives the system near $n_c$ into the insulator\,\cite{aniso_2002}.
As was pointed out in Ref.\,\cite{das_PRL_2006},
the MR data for in-plane fields, analyzed in Refs.\,\cite{shashkin_PRL_2001, vitkalov_PRL_2001},
is only approximately related to the spin susceptibility of {\em itinerant} electrons,
because it ignores the magnetic field dependence of $\chi$ even in the ``metallic'' conduction regime\,\cite{granada}.

\subsection{Spin polarization probed by the thermodynamic $\chi_T$ measurements}

\begin{figure}[h]
\includegraphics[width=240pt]{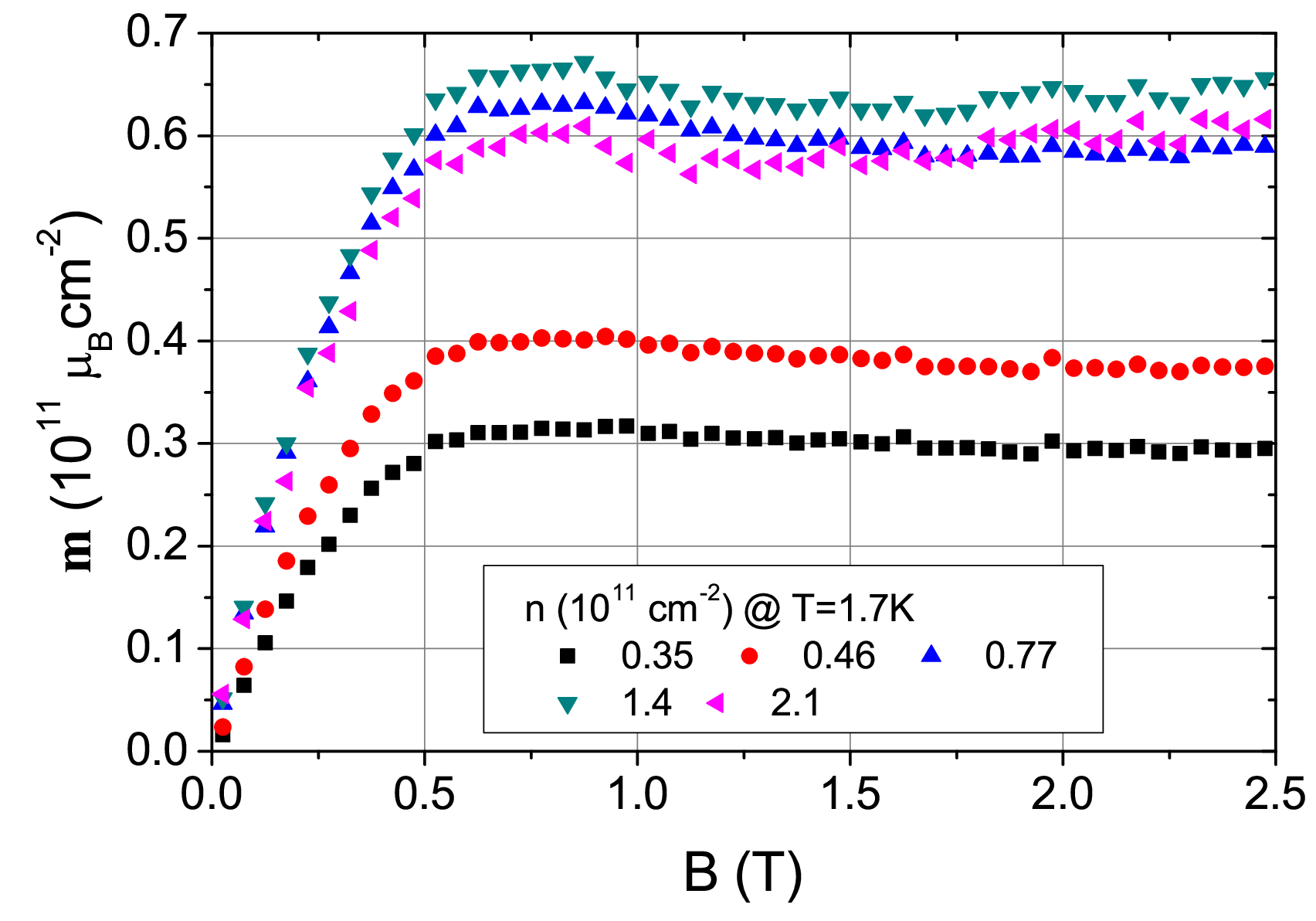}
\caption{(Color online)  Magnetic field dependence of the sample magnetization $M$ at $T$=1.7\,K
and several densities. $M$ is linear in $B$ in weak fields, and saturates
at values
below full polarization. Decrease of $M$ with increasing $B$ can be attributed to inaccuracy of subtracting the diamagnetic contribution to \dM{}\cite{reznikov_JETPL_2010}.
The data for a ``large'' sample.
}
\label{Fig:M_T}
\end{figure}

After
setting the upper bound on the degree of spin polarization of the itinerant electrons for $T\rightarrow 0$, we shall discuss the polarization of the whole system measured with recharging technique.

In the same way as \cT{} was obtained by integration of \dchi{}, we obtained $M(B)$ by integration of \dM{} at different magnetic fields and temperatures.
The results for $T=1.7$\,K are shown in Fig.\,\ref{Fig:M_T}.  The magnetic moment of the system as a whole grows linearly with the field, thus excluding zero-field spin polarization. Indeed, if it existed, it would lead to an offset: a finite $M$ at $B\rightarrow 0$.
At low densities below $n_c=8\times 10^{10}\,{\rm cm^{-2}}$  for the ``large'' high mobility sample, the magnetic field of 0.5\,T  almost polarizes the system. When the density increases above $n_c$, the magnetic moment at saturation stops to grow  with $n$, and even somewhat decreases.

At some temperature-dependent density, close but above $n_c$, sample magnetization starts to decrease with density: e.g. magnetization for $n=2.1\cdot 10^{11}\,{\rm cm^{-2}}$ is smaller than magnetization for $n=1.4\cdot 10^{11}\,{\rm cm^{-2}}$ This is consistent with the  non-monotonic density dependence of \cT{} shown in Fig.\,\ref{Fig:chi}.
There is also a slight decrease of magnetization with magnetic field above $B\approx 0.7$\,T.
We attribute it to inaccuracy of subtraction of the diamagnetic contribution: this contribution, being unimportant at weak magnetic fields, becomes important at strong, at which the paramagnetic contribution saturates.

Finally, we note that the
exponent $\alpha$ in the temperature dependence of \cT{} in Fig.\,\ref{Fig:chiT} shows no critical behavior at $n=n_c$:  it increases monotonically with decrease of the density.
In Refs.\,\cite{shashkin_PRL_2006, krav_SSC_2007} the  opposite conclusion was drawn from similar measurements of \cT{} by erroneously identifying the densities at which \cT{}
reaches maxima (see Fig.\,\ref{Fig:chi}) with the onset of full spin polarization, see Ref.~\cite{ReznikovCondmat04} for discussion.

\section{Conclusions}

To summarize, we found that:
\begin {enumerate}
\item
 Spin susceptibility $\chi^*$ of itinerant electrons in a strongly correlated 2D electron system in Si
inversion layer weakly depends on temperature over the range $0.05- 1$\,K.
\item
 This weak $\chi^*(T)$-dependence does not support non-FL predictions for the power-law divergence of $\chi^*$ as $T\rightarrow 0$.
\item
The temperature dependence of $\chi^*$ in  low magnetic fields  is consistent with the FL interaction corrections.
In contrast, the temperature dependence of $\chi^*$  in the magnetic field $B_\parallel>T/\mu_B$ seems somewhat stronger than the interaction correction
in ballistic regime predicts; on the other hand, it does not vanish, as expected for the interaction correction in diffusive regime.
\item
 The above inconsistencies might stem from the fact that the diffusive-ballistic crossover  in the susceptibility sets at temperature much lower
 than the conventional  $T =1/\tau$ value.
\item
Thermodynamic spin susceptibility $\chi_T$ is much larger than $\chi^*$, and its temperature dependence is much stronger.
We attribute this large susceptibility  to the localized spin droplets~\cite{teneh_PRL_2012}.
Their contribution to thermodynamics, and, particularly, to the spin susceptibility of the two-phase state is dominant;
it is magnetic field and temperature dependent and,  therefore, may also affect the  $\Delta\chi^*(T,B)$ dependence
observed in our experiments in  in-plane magnetic fields.
\item
The striking difference of the weak $\chi^*(T)$ and strong $\chi_T(T)$ dependences  is  evidence for a two-phase state
consisting of easily polarisable spin droplets
and an FL, even for the well conducting ``metallic'' regime $k_Fl \gg 1$.
\item
By analyzing both thermodynamic and transport data we  exclude the possibility of  a magnetic instability in Si 2DES
at any density above $5.3\times 10^{10}$\,cm$^{-2}$, including the MIT critical density $n=n_c$. The  complete spin polarization is absent
both in zero and low fields,  in the accessible range of temperatures (down to 100\,mK).
\end{enumerate}

We conclude that the FL type interaction corrections describe the measured $\chi^*(T)$ dependence for itinerant electrons reasonably well. The observed  deviations from the FL picture may be caused by the presence of the polarized spin-droplets, whose coexistence with the FL-state
is evidenced by the thermodynamic magnetization data.
In general, interacting electron  systems often have a tendency to phase separation in the vicinity of the  MIT\,\cite{basov_Sci_2007} or superconductor-insulator transition\,\cite{kornilov_PRB_2004, gerasimenko_PRB_2014}.
The concept of the 2D electron liquid as the two-phase
state\,\cite{morgun_PRB_2016, pudalov-ICSM-2016, pudalov-superstripes_JSNM_2016}
is capable of qualitatively explaining thermodynamic magnetization data.
The quantitative description of $\chi_T(T)$ and $\chi^*(T)$ dependences, and, particularly,  understanding the crossover regime in $\chi^*(T)$, however, requires further studies.

\begin{acknowledgements}
Authors are grateful to B.~L.~Al'tshuler, A.~V.~Chubukov, D.~V.~Efremov, A.~M.~Finkel'stein,  D.~L.~Maslov, O.~Sushkov,  and A.~Yu.~Zyuzin  for
discussions,  and to N.~Teneh for his assistance in thermodynamic  magnetization measurements. VMP is  supported by  RFBR No18-02-01013 (measurements) and by Russian Ministry of Education and Science (Project 2017-14-588-0007-011), ISB is supported by  the Alexander von Humboldt Foundation,   MR acknowledges Israeli Science Foundation, and
MG --  the NSF Grant DMR 0077825.
\end{acknowledgements}

\appendix
\section{Interaction corrections to $\chi^*(B,T)$
\label{App1}}

In this appendix we discuss briefly theoretical background for the interaction corrections to the spin susceptibility $\chi^*(B,T)$ in the diffusive and ballistic regimes.

\subsection{Diffusive regime}

The original Al'tshuler-Zyuzin  result\,\cite{altshuler82} predicts the following behavior of the spin susceptibility of  a disordered 2D electron system in the diffusive regime:
\begin{equation}
\Delta\chi(T) \propto - \begin{cases}
\ln (T\tau), & \quad B_\parallel=0, \\
\ln ( g\mu_B B_\parallel\tau), & \quad g\mu_B B_\parallel \gg T .
\end{cases}
\label{Altshuler}
\end{equation}
Extension of the original result of Ref.\,\cite{altshuler82} to an arbitrary value of the ratio $g\mu_B B_\parallel/T$ and an arbitrary number of valleys, $n_v$ can be done under the following assumptions: (i) the system is in the metallic regime, i.e. zero-field conductivity is large, $\sigma_{xx}(0)  \gg e^2/h$; (ii) the temperature is low enough: $(1 + \gamma_2)T \tau \ll 1$; and (iii) the parallel magnetic field is not too strong  $(1 + \gamma_2)g\mu_BB_\parallel \tau \ll 1$.
 Here, we remind that parameter $\gamma_2$ determines FL renormalization of  $g$-factor, $g^*=(1+\gamma_2) g$, and can be expressed through the temperature-independent FL parameter ${F_0^\sigma}$: $\gamma_2= -{F_0^\sigma}/{(1+F_0^\sigma)}$. For the representative density $2\times 10^{11}$cm$^{-2}$ the parameter $\gamma_2\approx 0.6$.
Following the approach of Refs.\,\cite{Burmistrov2008,Burmistrov2017}, one can find:
\begin{equation}
\frac{\chi(B_\parallel)}{\chi(0)}= 1-\frac{1}{\sigma_{xx}(0)}\frac{n_v^2(1+\gamma_2)}{\pi^2} {\cal{F}}\left( \frac{g\mu_B B_\parallel}{2\pi T}\right).
\label{eq:1}
\end{equation}
Here $(0)$ stands for zero field,  and $\sigma_{xx}(0)$
is in units of $e^2/\hbar$:
\begin{equation}
\frac{\chi(0)}{\chi_P} =\left[1+ \frac{1}{\sigma_{xx}(0)}\frac{n_v^2\gamma_2}{\pi^2}\left(\ln\frac{1}{2\pi T\tau} -\psi(1)  \right) \right] ,
\label{eq:B0}
\end{equation}
and the function ${\cal{F}}$ is expressed through the Euler di-gamma function:
\begin{equation}
{\cal{F}}(h)=
\textrm{Re}\, \Bigl [
\psi\bigl (1+i(1+\gamma_2)h\bigr )- \frac{\psi(1+ih)+\gamma_2\psi(1)}{1+\gamma_2}
\Bigr ] .
\end{equation}
We emphasize that Eqs.~\ref{eq:1} and \ref{eq:B0} are derived to the lowest order
in $1/\sigma_{xx}(0)$ (in the dimensionless units, as stated above).

At weak fields and not too low temperatures, $(1+\gamma_2)g\mu_B B_\parallel/(2 \pi T) \ll 1$, one can expand the function $\mathcal{F}(h)$ to the second order in $h$, and find:
\begin{equation}
\frac{\chi(B_\parallel)}{\chi(0)}= 1-\frac{1}{\sigma_{xx}} \frac{\zeta(3)}{\pi^2}n_v^2[(1+\gamma_2)^3-1]\left(\frac{g\mu_B B_\parallel}{2\pi T}  \right)^2 .
\label{eq:lowB}
\end{equation}

In the opposite limit of low temperatures, $2 \pi T \ll (1+\gamma_2) g\mu_B B_\parallel$, we obtain the following asymptotic behavior:
\begin{gather}
\frac{\chi(B_\parallel,T)}{\chi(0)}=  1-\frac{1}{\sigma_{xx}}\frac{n_v^2\gamma_2}{\pi^2} \Biggl [\ln\frac{g\mu_B B_\parallel}{2\pi T} -\psi(1)
\notag \\
+ \frac{1+\gamma_2}{\gamma_2}\ln(1+\gamma_2)  \Biggr] .
\label{eq:highB}
\end{gather}
This result implies
\begin{equation}\label{A7}
\frac{\chi(B_\parallel) -\chi_P}{\chi_P} \approx  -\frac{1}{\sigma_{xx}}\frac{n_v^2\gamma_2}{\pi^2} \ln(g\mu_B B_\parallel \tau),
\end{equation}
that is perfectly consistent with the well-known
result, Eq.\,(\ref{Altshuler})\,\cite{altshuler82}.

\begin{figure}[h]
\begin{center}
\includegraphics[width=200pt]{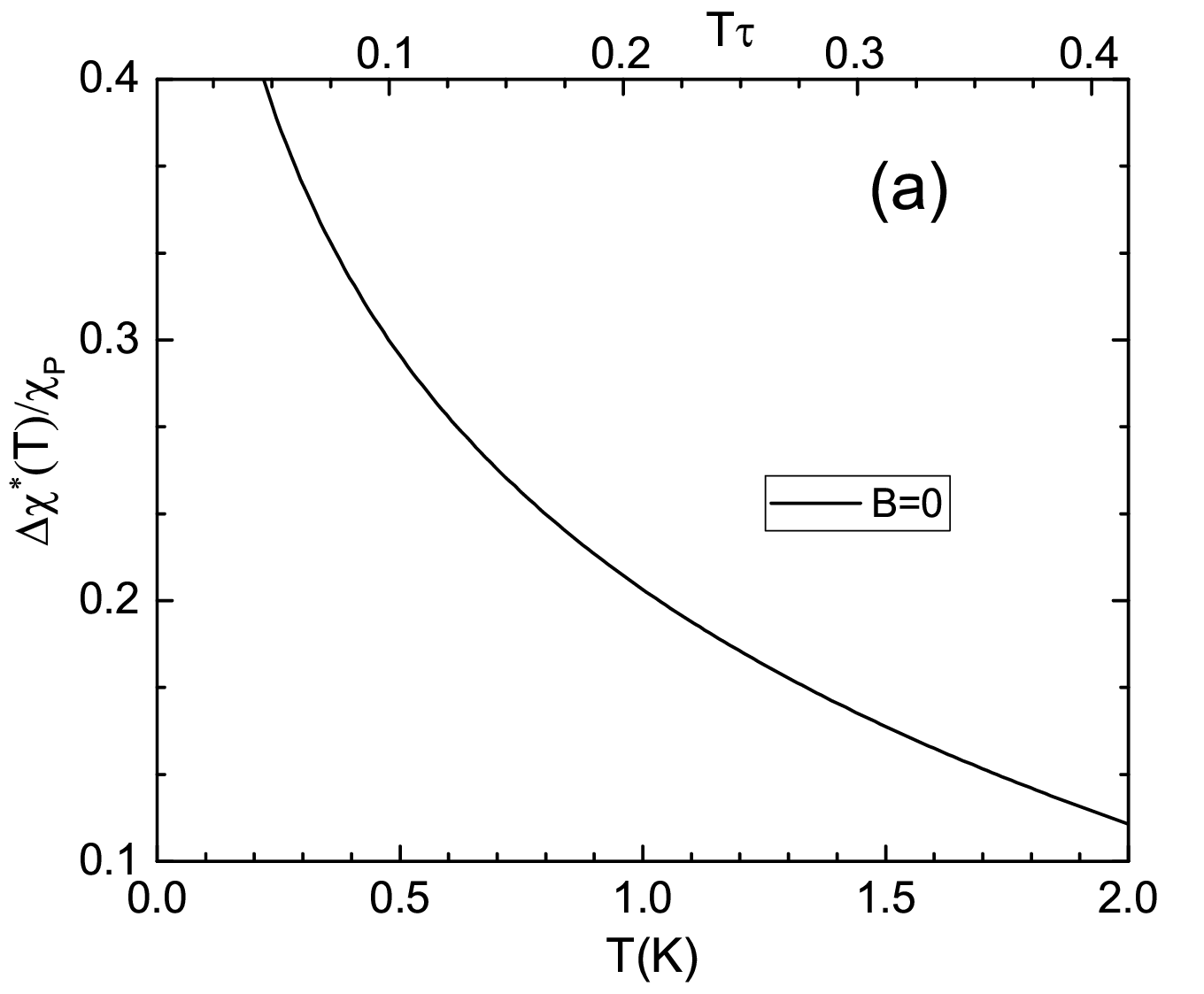}
\includegraphics[width=200pt]{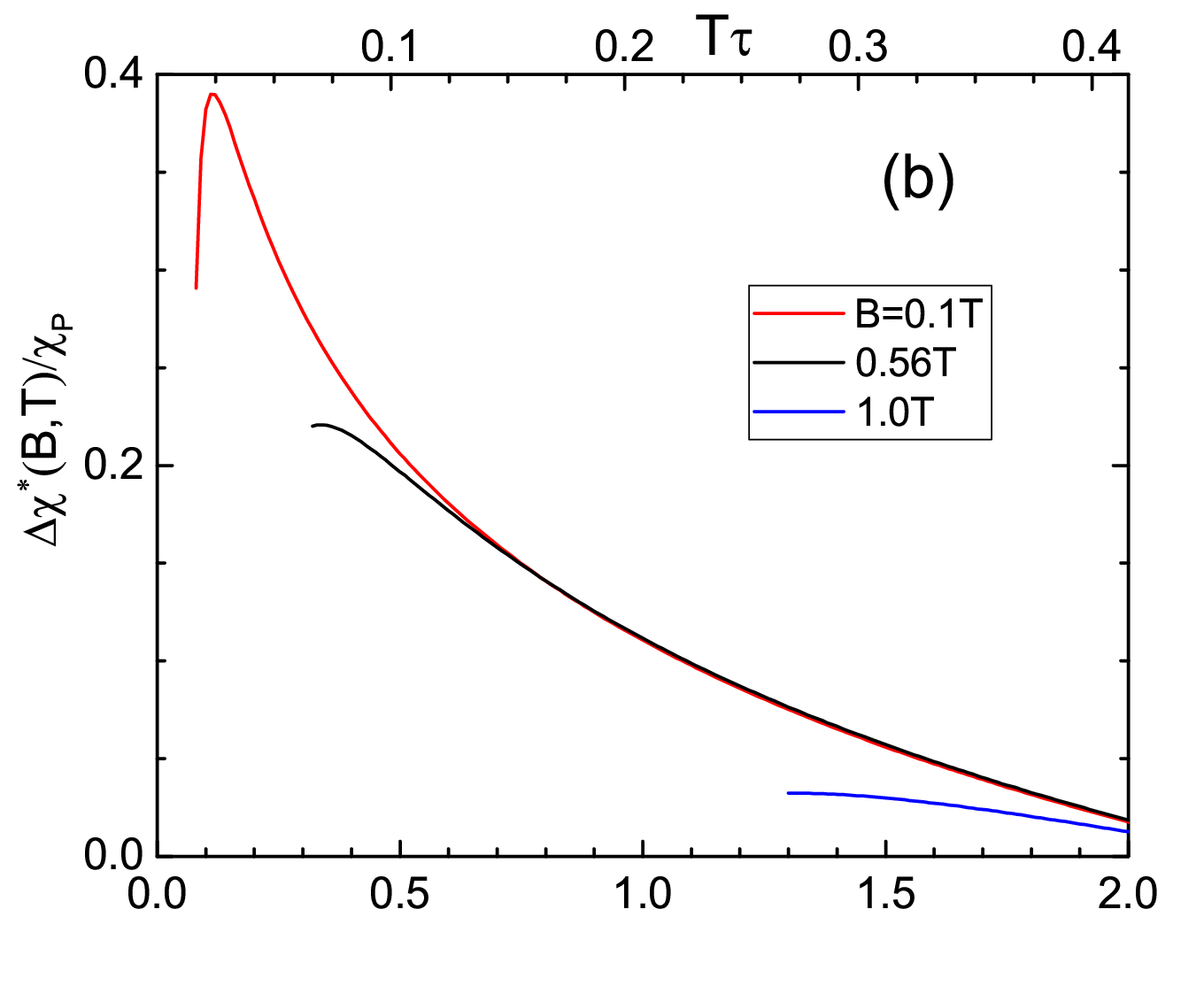}
\caption{
Temperature dependence of the diffusive interaction correction $\Delta\chi^*(T)$ for parameters corresponding to Fig.\,\ref{fig4}(b): $n=2.08\times 10^{11}$cm$^{-2}$, $\tau=1.59$ps, $F_0^\sigma=-0.38$, $\gamma_2=0.61$, ``small'' sample. (a) calculated from Eq.~(\ref{eq:B0}) for $B=0$; temperature interval is cut by requirements $\Delta\chi^*(T)/\chi_P <0.4$ ($T\geq 0.05$K),  and $T\tau <0.42$ ($T<2$\,K).
(b)  Calculated from Eq.~(\ref{eq:lowB}) for   $B=0.1$,   0.56\, and 1.0\,T.
For the panel (b), 
temperature is limited from below by the requirement $(1+\gamma_2)g\mu_B B_\parallel/(2 \pi T) < 1$, and from above,  by $T\tau <0.42$ ($T<2$\,K).
Vertical position is chosen to obtain vanishing  $\Delta\chi^*(T)$  for $T\tau = 1$.
}
\label{fig:diffusive_Burmistrov}
\end{center}
\end{figure}

In Figure~\ref{fig:diffusive_Burmistrov} we show the $\Delta\chi(B_\parallel,T)$ dependences calculated from  Eq.\,(\ref{eq:B0}) for $B_\parallel=0$ and
from Eqs.\,(\ref{eq:lowB}), and (\ref{eq:highB}) for  $B_\parallel=0.1$,  $0.5$\,T and 1\,T.
Note, that $(1+\gamma_2)g\mu_BB_\parallel\tau$ becomes unity  at $B\sim 2.3$\,T ($\tau$ is indicated in the upper horizontal scale of Figs.\,\ref{fig4}(a)-(d), and are also given in the captions to Fig.\,\ref{fig:diffusive_Burmistrov}), so that assumption (iii) of its smallness  is violated for  experimental parameters of Fig.\,\ref{fig4}d, and to some extent, of Fig.\,\ref{fig4}(c).  In such high fields,  Eq.\,(\ref{A7}), according to which $\chi^*$ is temperature independent, is more appropriate.

The logarithmic increase of $\Delta\chi^*$ with lowering temperature comes from $\chi(0)$, Eq.\,(\ref{eq:lowB}).  The maximum and the further decrease seen at very low temperatures (and not observed in the experiment) for $B_\parallel=0.1$\,T comes from the field-dependent correction in Eqs.\,(\ref{eq:lowB}), and (\ref{eq:highB}).

 \subsection{Ballistic regime}

\begin{figure}[h]
\begin{center}
\includegraphics[width=200pt]{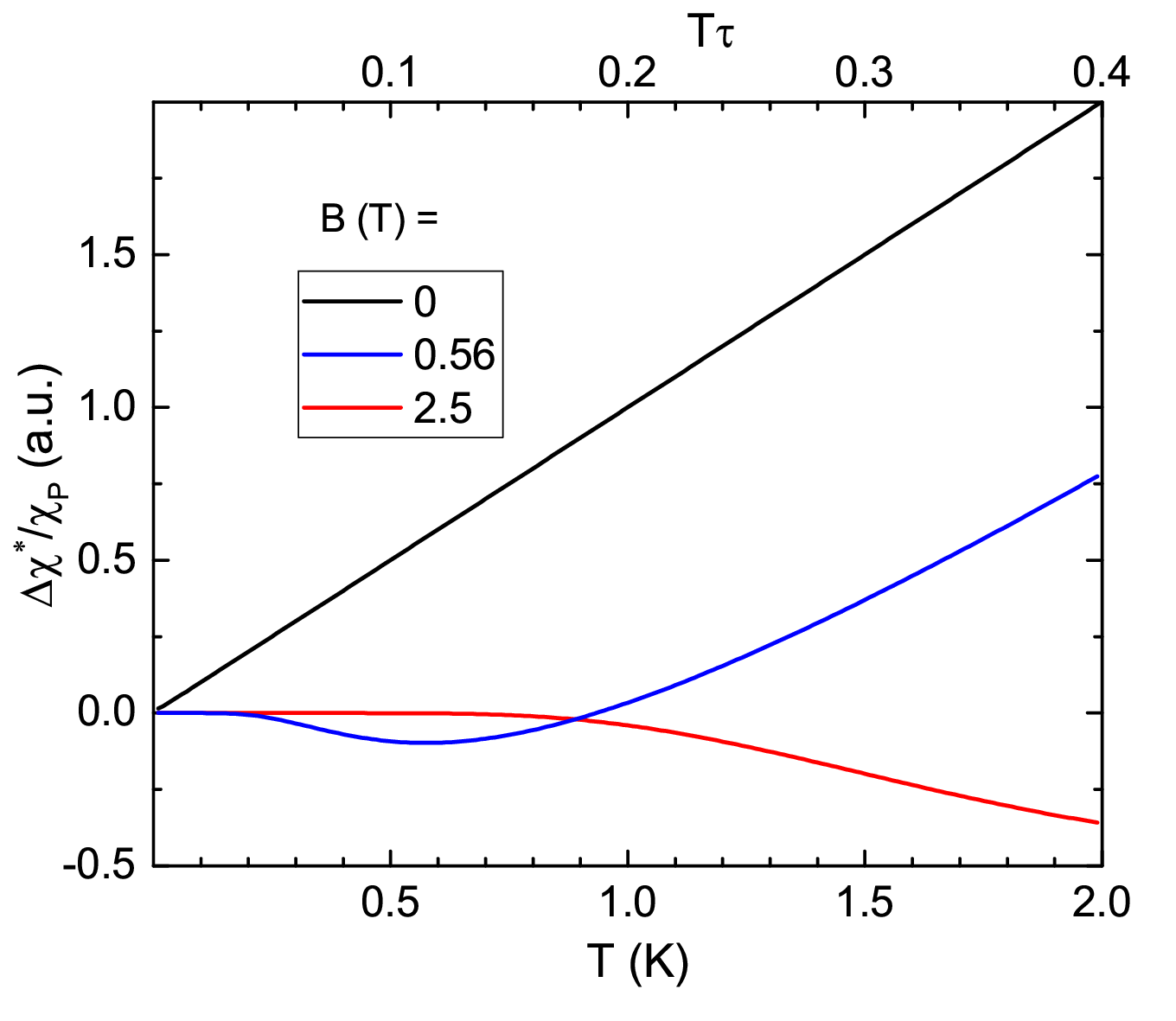}
\caption{ Ballistic correction $\Delta\chi^*(T)/\chi_P$ calculated from Eq.~(\ref{eq:efremov}) of \cite{betouras-efremov_PRB_2005}
for parameters corresponding to Fig.\,\ref{fig4}(a) -- 4(c):
$n=2.08\times 10^{11}$cm$^{-2}$, $\tau=1.59$ps, the ``small'' sample,
at $B=0$, 0.56, and 2.5\,T. Vertical scale is the same for all three curves. Upper scale shows temperature in units of $T\tau$.
}
\label{fig:ballistic efremov}
\end{center}
\end{figure}

For completeness, we remind of the result for the interaction correction to the spin susceptibility $\Delta\chi^*(T) = \chi^*(T)-\chi^*(0)$ in the ballistic regime, $T\tau \gg1$.
At zero field, $\Delta\chi^*(T)$ for a 2D system increases approximately linear with $T$  (see Eq.\,(\ref{chi}))
and\,\cite{chitov-millis, chubukov-maslov_PRB_2003, galitski-das_PRB_2005, betouras-efremov_PRB_2005, chubukov-maslov_PRL_2005, belitz97}). This linear in $T$ dependence appears in the second order in  the interaction corrections due to nonanalytic behavior of the polarization operator at $q=2k_F$.
In Ref.\,\cite{betouras-efremov_PRB_2005}, the correction to the spin susceptibility was computed in the presence of a non-zero in-plane magnetic field. The result is the following:
\be\label{eq:efremov}
\Delta \chi^*(T) = \chi_P A^2(\pi) T f\bigl (g\mu_B B_\parallel/T\bigr ),
\ee
where
\be
f(x)=\frac{x}{\sinh^2(x)}\bigl[\sinh(2x)-x \bigr],
\ee
and $A\sim 1$.
To compare this prediction with the experimental data in Figs.~4(a)\,--\,4(c) we show  in Fig.\,\ref{fig:ballistic efremov} $\Delta\chi^*(T)= \chi^*(T)-\chi^*(0)$ dependence
calculated from Eq.~(\ref{eq:efremov})  for $B_\parallel=0$, 0.56, and 2.5T.
Though there is a certain similarity, the whole explored temperature range formally belongs to the diffusive interaction regime
$T\tau \ll 1$
and the applicability of the ballistic corrections (Fig.\,\ref{fig:ballistic efremov})  to the data  Fig.\,\ref{fig4} is questionable.

\section{Experimental determination of the carrier density}
\label{sec:App_density}
The {\it full carrier density} $n$ in the 2D channel of  Si-MOSFETs   was determined from the ShdH oscillations and from the Hall effect measurements.
Both methods are applicable down to the critical density  $n_c \approx 1\times 10^{11}$\,cm$^{-2}$,
 and give the same results with accuracy of a few percent \cite{shashkin_PRL_2001}.
Within the same accuracy, these results also agree with
the charge in the gated Si-MOSFET directly measured by recharging or capacitive-type
techniques \cite{pudalov-charge_JETPL_1984}:\\
$$n = \frac{C}{S e} (V_g - V_{th}).$$
Here $V_{th}$ is the gate voltage at which density extrapolates to zero and $S$ is the sample area.
The calculated value of capacitance $C$  agrees with the directly measured capacitance
of the sample.

{\it Density of itinerant and localized electrons.\\}
 We presume the itinerant electrons to be those having a $ps$-scale relaxation time and therefore providing a dominant contribution to the DC conductance and ShdH.
In contrast, the localized carriers are trapped for a much longer time, and their contribution to the conductivity is negligible.
Both ShdH and Hall effect measurements give results close to the  full electron density. At high densities deep in
the metallic phase and at low magnetic fields in a uniform system all electrons are itinerant,
and $n_i = n$.

In a nonuniform system, ShdH and Hall effect also measure the
 full electron density.
ShdH counts the number of filled Landau
levels, without distinguishing between localized and extended states, an extreme example
is the QHE: there might be a few extended states in the middle of the Landau levels, but the density
extracted from the oscillations period  is just the
 full one, $n$. In the QHE case, the
Hall effect is stepwise;  the width of the steps is an indication
for the number of localized states in this regime. The density of localized states is known to oscillate
 with  filling factor or Fermi energy position relative to the Landau levels. We are not aware of a
straight-forward way to infer the number of localized electrons in low magnetic fields close
to the MI transition. Our measurements of the number of the easily-polarized spins can be
an indication.

Anyhow, the knowledge of $n_i$ at different scales is not
required for the  subject of this paper: namely, the susceptibility and its properties
with  $n$ and $T$ being the governing parameters.

\section{Fitting of the  SdH oscillations
\label{sec:Appendix_2}}
The  magnetooscillations in the {\em noninteracting} Fermi gas are usually fitted \cite{gm, pudalov_PRB_2014} by the  Lifshitz-Kosevich (LK) formula,
adapted for the 2D case and valid for
a small amplitude of oscillations $\delta\rho/\rho \ll 1$ \cite{SdH} :
\begin{eqnarray}
\frac{\delta\rho_{xx}}{\rho_0}&=& +2\frac{\Delta D}{D}=\sum_i
A_i^{\rm LK}\cos\left[ \pi i\left(\frac{\hbar c \pi n }{e B_\perp}
-1\right)\right]
Z_i^s Z_i^v.
\nonumber\\
A_i^{\rm LK}&=&4\exp\left(-\frac{2\pi^2 i k_B
T_D}{\hbar \omega_c}\right) \frac{2\pi^2 i k_BT/\hbar \omega_c}{\sinh\left({2\pi^2 i
k_BT}/{\hbar \omega_c}\right)}
\label{SdHbyLK}
\end{eqnarray}
Here  $\omega_c=eB_\perp/m^*c$ is the cyclotron frequency, $D$ is the 2D density of states,
$T_D$ is the Dingle temperature,
and the Zeeman- and valley- splitting terms are:

\begin{equation}
Z_i^s=\cos \left[\pi i\frac{ \hbar \pi P c n}{e B_{\perp}}\right], \quad
Z_i^v=\cos\left[ \pi i \frac{\Delta_V}{\hbar\omega_c}\right],
\label{zeeman_splitting}
\end{equation}
 where $P=(n_\uparrow - n_\downarrow)/n$ is the spin polarization,
 $n_\uparrow$ ($n_\downarrow$) stand for the density of spin-up (spin-down) electrons, and $\Delta_V$ is the valley splitting\,\cite{ando_review}.

In the absence of in-plane field, the Zeeman term reduces to the field-independent factor
$$
\cos\left(\pi i\frac{\Delta_Z}{\hbar\omega_c}\right)=\cos\left(\pi i\frac{g^*m^*}{2m_e}\right).
$$

In the case of {\em interacting} electrons, the temperature dependence of the oscillation amplitude changes\,\cite{martin_prb_2003, adamov_prb_2006, pudalov_PRB_2014}.
The modified  amplitude of the oscillations is a monotonic function of $B_\perp$ \cite{pudalov_PRB_2014}
and its exact shape does not affect fitting in the narrow field interval in the vicinity of  the nodes (Fig.\,\ref{fig2}).

\section{Domain of measurements of the SdH oscillations close to $n_c$ \label{App3}}

\begin{figure}[h]
\begin{center}
\includegraphics[width=170pt]{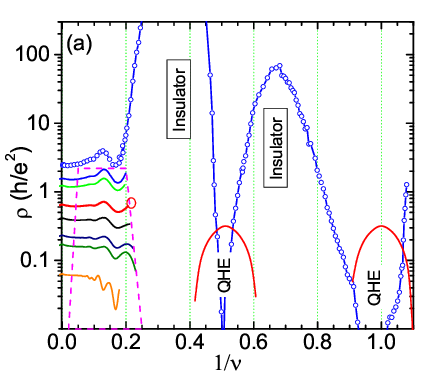}
\includegraphics[width=180pt]{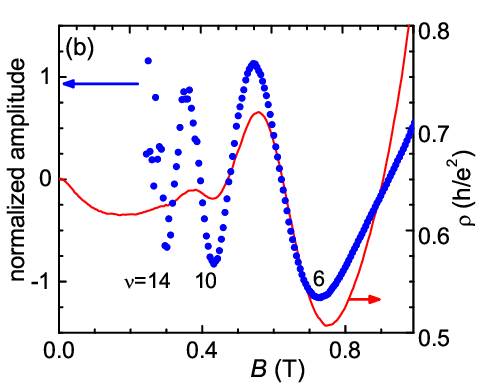}
\caption{
Evolution of the ShdH oscillations
at different densities close to the critical density  of MIT for  the ``small'' sample  ($n_c \approx 0.97\times 10^{11}$cm$^{-2}$).
Empty circles  in panel (a) show the $\rho_{xx}$ oscillations  at $n\approx 0.97\times 10^{11}$cm$^{-2}$ which in high fields are transformed into the reentrant
QHE-insulator transitions \protect\cite{pudalov_PRB_1992}. Dashed line  confines the region of our weak field ShdH measurements in
Ref~\protect\cite{spin-polarized}. The oscillatory curves in the confined regions are for densities (from top to bottom) 0.98, 0.99, 1.04, 1.10, 1.20, 1.30, and 1.98 in units of $10^{11}$cm$^{-2}$.
(b) Right axis: expanded view of one of the $\rho_{xx}(B)$ curves  at $n=1.04\times
10^{11}$cm$^{-2}$ [marked with a circle on panel (a)]; left axis:  the MR oscillatory component  after subtraction the monotonic magnetoresistance and normalization by the amplitude of the first
harmonic $P_1(B_\perp)$  \protect\cite{gm}.
}
\label{fig:phase_digram}
\end{center}
\end{figure}

Figure\,\ref{fig:phase_digram}(a) illustrates
the parameter  range, in which the quantum oscillations were measured in the vicinity of the insulating regime. For a density close to  $n_c$, as the perpendicular field increases, the 2D system experiences reentrant QHE-insulator transition (see Refs.\,\cite{pudalov_PRB_1992, termination94}).
In weaker fields, however, the oscillations look quite ordinary, enabling their conventional analysis, as we described above.

The oscillation amplitude in  Fig.\,\ref{fig:phase_digram}(b) is damped by a factor of  $\sim 1/50$ in the field of $B_\perp=0.4$\,T; converting this  ``Dingle factor'' to the quantum (all angle) scattering time $\tau_q=\hbar/2\pi k_BT_D$, we conclude that  the 2D system approaches  critical density with rather high $\omega_c\tau_q \sim 1$ value in $B_\perp=0.4T$.
On the other hand, treating the DC conduction with  Drude formula, we find the momentum relaxation time to be several times smaller.
The fact that $\tau_q$  exceeds the momentum scattering time $\tau$  when
$n$ approaches $n_c$  is puzzling, though  well known.
Its discussion  is beyond the framework of the current paper; we only note that such unusual relation between the scattering times
indicates failure of the homogeneous FL state and may be understood within
the concept of the two-phase state \cite{teneh_PRL_2012, morgun_PRB_2016}.

\begin{thebibliography}{99}
\bibitem{note1} The parameter
$r_s = 2.63\times (10^{12}{\rm cm}^{-2}/n)^{1/2}$
which characterizes the electron-electron interactions,
is the  ratio of the Coulomb interaction energy $E_{ee}$ to the Fermi
energy $E_F$. The bare values of the Fermi energy $E_F$, the effective mass $m_b=0.19$,
and the dielectric constant $\kappa =7.8$ at the
Si-SiO$_2$ interface are used in its definition.
\end{thebibliography}
\end{document}